\UseRawInputEncoding
\documentclass[
 reprint,
 aps]{revtex4-2}
\usepackage{graphicx}
\usepackage{stackengine}
\usepackage{epsfig}
\usepackage[caption=false]{subfig}
\usepackage{subfloat}
\usepackage{amssymb}
\usepackage{amsmath}

\usepackage{cleveref}

\usepackage[siunitx, europeanresistors]{circuitikz}
\usepackage{tikz}
\usetikzlibrary{fit}
\usetikzlibrary{arrows}

\begin{document}

\title{Feedback stabilization of the resonant frequency\\ in tunable microwave cavities with single-photon occupancy}

\author{S. Kanhirathingal}
\email{Sisira.Kanhirathingal.GR@dartmouth.edu}
\affiliation{Department of Physics and Astronomy, Dartmouth College, Hanover, New Hampshire 03755, USA.}
\author{B. Thyagarajan}
\affiliation{Department of Physics and Astronomy, Dartmouth College, Hanover, New Hampshire 03755, USA.}
\author{B. L. Brock}
\thanks{Department of Applied Physics, Yale University, New Haven, Connecticut, USA.}
\author{Juliang Li}
\thanks{Argonne National Laboratory, Argonne, Illinois, USA.}
\author{E. Jeffrey}
\thanks{Google Research, Santa Barbara, California, USA.}
\author{M. P. Blencowe}
\affiliation{Department of Physics and Astronomy, Dartmouth College, Hanover, New Hampshire 03755, USA.}
\author{J. Y. Mutus}
\thanks{Rigetti Computing, Berkeley, California, USA.}
\thanks{Google Research, Santa Barbara, California, USA.}
\author{A. J. Rimberg}
\email{Alexander.J.Rimberg@dartmouth.edu}
\affiliation{Department of Physics and Astronomy, Dartmouth College, Hanover, New Hampshire 03755, USA.}
\date{\today}

\begin{abstract}
We successfully demonstrate low-frequency noise suppression in the resonant frequency fluctuations of a cavity-embedded Cooper pair transistor (cCPT) driven at single-photon occupancy. In particular, we report a reduction in the resonant frequency fluctuations caused by the internal charge noise over a bandwidth of $\sim$1.4 kHz when the cavity is driven at an average photon number $n=10$, and a bandwidth of 11 Hz for average $n=1$. The gate-dependent tunability of the cCPT allows us to implement a feedback-scheme derived from the Pound-Drever-Hall locking technique. This reduces fluctuations due to intrinsic charge-noise, which otherwise interferes with the cCPT's operation as a near quantum-limited electrometer. Our technique can be generalized to achieve frequency stabilization in tunable microwave resonators that play a vital role in today's quantum computing architectures, thereby moderating the limitations in detection caused by the intrinsic $1/f$-noise on such circuit devices. The work discusses the various aspects relating to the operation of a fully functional feedback loop down to the single-photon level. 
\end{abstract}
\maketitle
\section{Introduction}
\label{intro}
The existence of two-level-system induced $1/f$-noise is well-known to limit the efficiency and sensitivity of devices across a breadth of applications -- ranging from the semiconductor industry, to the emerging field of quantum computing processors~\cite{muller2019towards}. Understanding its microscopic origin~\cite{faoro2015interacting,anderson1972anomalous,schlor2019correlating} and exploring different approaches to suppress this noise is a crucial step towards the realization of high coherence superconducting quantum circuits~\cite{paladino20141,burnett2019decoherence,anton2012pure,lu2021quantum}, ultra-sensitive electrometry/magnetometry ~\cite{brock2021fast,zorin1996background,gustafsson2013thermal,zimmerli1992noise,wellstood1987low,hatridge2011dispersive,levenson2013dispersive}, and other studies more fundamental in nature~\cite{cattiaux2021macroscopic,fefferman2008acoustic}. 

Many approaches to reduce low-frequency noise focus on the elimination of two-level defects on the hosts, during fabrication and post-processing~\cite{de2018suppression,nguyen2019high,kumar2016origin, earnest2018substrate,muller2019towards,mcrae2020materials}. Besides often being a cumbersome task that can also sometimes be expensive to implement, some of these methods can cost anharmonicity of energy levels, which are critical for the performance of qubits~\cite{koch2007charge}. Such systems can therefore profoundly benefit from the real-time detection and suppression of $1/f$-noise while performing measurements, thence significantly improving their performance~\cite{nakajima2021real,tian2007correcting,muck1994investigation}.

One system which displays strong charge sensing properties at very low pump powers, and which suffers from reduced sensitivity due to intrinsically induced low-frequency noise, is the cavity-embedded Cooper pair transistor(cCPT)~\cite{brock2021nonlinear}. The cCPT is a nonlinear charge- and flux-tunable microwave cavity and its complete noise characterization presented in Ref. \cite{brock2021nonlinear} addresses the role of the intrinsic noise in charge/flux bias leading to resonant frequency fluctuations, especially in regions where the cCPT can operate as a highly sensitive electrometer/magnetometer. By singling out bias regions where the cCPT is maximally sensitive to charge/flux fluctuations, measurements detected typical charge and flux noise spectral densities of the form $S_{\mathrm{qq}} \propto 1/f\; \text{e}^2/\text{Hz}$, and $S_{\mathrm{\Phi \Phi}} \propto \sqrt{1/f}\;\Phi_0^2/\text{Hz}$, respectively. The magnitude of these resonant frequency fluctuations at some bias points is of the order of the cavity linewidth, shifting the carrier signal away from the cavity resonance during the course of a measurement. As a result, although the cCPT is capable of achieving quantum-limited electrometry at very low pump powers~\cite{kanhirathingal2021charge}, the observed charge sensitivity is nearly three times worse than the theoretical predictions~\cite{brock2021fast}.

This work reports a reduction of these frequency fluctuations induced by the intrinsic charge/flux noise on the cCPT. Such a study is of two-fold importance to the general circuit-QED audience. Firstly, in many ways the cCPT mimics the resonant tunability and readout scheme generally adopted in quantum computing architectures~\cite{jkelly}, while working with a simpler circuit system. The basic structure consists of a quarter-wavelength superconducting microwave resonator (in a coplanar waveguide geometry), with non-linear tunability introduced via a Cooper pair transistor (CPT) formed using two Josephson junctions in series. Dispersive reflection measurements of the resonator via capacitive coupling to a pump/probe transmission line enable readout of the system state. Similar to the devices mentioned above, the cCPT is exposed to low-frequency charge noise due to charge traps nearby the CPT island, as well as to flux noise originating from the unpaired surface spins coupling to the SQUID loop. As the cCPT is specifically designed to be a highly sensitive electrometer/magnetometer, it is an ideal candidate for understanding and suppressing the associated effects of such $1/f$-noise commonly found in these devices. Secondly, stabilizing the resonant frequency fluctuations can potentially elevate the cCPT into a superior charge sensing regime compared to previously reported results for the same cCPT device~\cite{brock2021fast}. Ultrasensitive electrometry can aid in the realization of a macroscopic optomechanical system in the single photon-phonon strong coupling regime as proposed in \cite{rimberg-cavity-cooper-2014,heikkila2014enhancing,nunnenkamp-single-photon-2011}. Furthermore, stabilizing against charge fluctuations can provide controllable access to the neighborhood of the Kerr-sourced bifurcation point of the cCPT, where the charge sensitivity undergoes a steep increase in magnitude~\cite{tosi2019design,laflamme2011quantum,brockthesis}.

The scheme to achieve the suppression of intrinsic bias-noise follows the well-established technique of Pound-Drever-Hall (PDH) locking, extensively used in laser optics to stabilize laser sources during cavity reflection measurements~\cite{black2001introduction}. Studies reporting the successful tracking of the resonant frequency fluctuations in superconducting microwave resonators utilizing this technique are also available in the literature~\cite{lindstrom2011pound,niepce2021stability, de2018suppression}. By carefully calibrating the circuit at each stage to provide maximum signal-to-noise ratio (SNR), we suppress intrinsic $1/f$-noise in the resonant frequency fluctuations over a bandwidth of 10 Hz, while driving the cavity at an average of merely a single photon. When the average photon number in the cavity is increased to $n=10$, this bandwidth increases to 1.4 kHz.

In the conventional approach to Pound-locking in microwave cavities, we utilize an error signal to correct the drive frequency such that it continuously tracks the fluctuating resonance. Some of the underlying factors leading to these resonant fluctuations include the dielectric losses due to the superconducting cavity's direct coupling to its immediate environment~\cite{burnett2014evidence}, and radiation noise leading to quasiparticle poisoning in the CPT~\cite{barends2011minimizing}. However, in general, the measured fluctuations follow a $1/f$-behavior as mentioned before, and are believed to emerge from two-level system (TLS) defects coupling through various channels into the cavity~\cite{anton2012pure,schlor2019correlating}. In the case of the cCPT and similar tunable microwave cavities, the dominant sources of these fluctuations are $1/f$-charge and flux noise coupling to the  resonant frequency via its tunability. 
Hence, when the cCPT is tuned to regions of maximum charge/flux sensitivity, this also results in the parametric coupling of unwanted electrical and magnetic fluctuations to the microwave cavity, leading to increased resonant frequency fluctuations. Locking to a stable reference thus results in a more stable resonant frequency of the cavity, significantly improving quantum sensing in these devices.

The paper layout is as follows. First, in Sec. \ref{concept}, we describe the basic circuit scheme that suppresses the resonant frequency fluctuations caused due to intrinsic charge/flux noise in tunable microwave cavities, along with a theoretical model using cavity field operators. Next, in Sec. \ref{ccpt_sec}, we discuss this scheme for the specific case of the cCPT, with particular consideration given to its Kerr-nonlinearity, as well as to the two-dimensional parameter space spanned by gate and flux tunability. We next provide the actual experimental setup in Sec. \ref{experiment}, discussing in detail the series of steps to maximize the SNR at the single-photon level. Following this, we report the results proving resonant frequency stabilization under feedback locking in Sec. \ref{results}.
We also provide insights into the applications and empirical limitations of the technique. Finally, in Sec. \ref{conclusion}, we present a summary of the results.
\section{Concept}
\label{concept}

\begin{figure}[thb]
\centering
\includegraphics[width=0.48\textwidth]{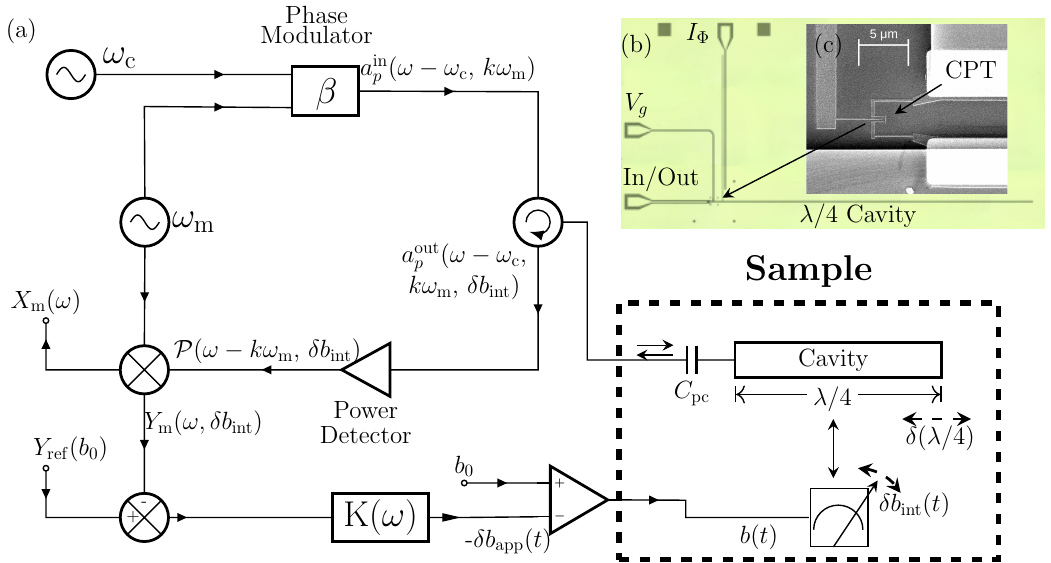}
\caption{\label{basic_circuit_scheme} (a) Basic feedback-based circuit scheme to stabilize cavity resonant frequency in the presence of intrinsic bias fluctuations $\delta b_{\mathrm{int}}(t)$. The phase-modulated input signal encodes the magnitude of bias fluctuations after reflection from the cavity. By continuously tracking and correcting for the fluctuations in the component of $\mathcal{P}(\omega-k\omega_\mathrm{m},\delta b_{\mathrm{int}})$ oscillating at frequency $\omega_\mathrm{m}$, we stabilize the resonance via an applied $\delta b_{\mathrm{app}}(t)$. (b) Sample image of the cCPT which is used to demonstrate resonant frequency stabilization in tunable microwave cavities. (c) Image of the Cooper pair transistor (CPT) that adds tunability to the $\lambda$/4 cavity. Detailed images and the experimental characterization of this device are reported by Brock \textit{et al.}~\cite{brock2021nonlinear}.}
\end{figure}

We will begin with a tunable cavity at resonance $\omega_0(b)$, displaying a linear reflection coefficient $S_{11}(\Delta)$, with tunability induced via parameter $b$, and detuning defined by $\Delta = \omega- \omega_0(b)$. The cavity undergoes resonant fluctuations due to undesired coupling with other systems in its environment. Let us assume that these fluctuations are dominated at any time by the intrinsic fluctuations in the bias parameter $b(t) = b_0 + \delta b_{\mathrm{int}}(t)$. The exact origins of these fluctuations are not of relevance in the current work. As discussed in Sec. \ref{intro}, we are especially interested in low-frequency noise where the power spectral density (PSD) of the bias noise, given by $S_{bb} (\omega)$, is predominantly $1/f$ in nature. As detailed below, Fig. \ref{basic_circuit_scheme}(a) then provides a feedback-based scheme to stabilize the resonant frequency fluctuations by effectively decoupling the low-frequency bias fluctuations from the cavity. 

The dashed box in Fig. \ref{basic_circuit_scheme}(a) represents our sample, containing a quarter-wave microwave resonator tunable via the parameter $b$. The cavity undergoes reflection measurements and is connected to the external drive-pump/ measurement-probe transmission line via a coupling capacitor $C_{\mathrm{pc}}$. Due to the intrinsic noise $\delta b_{\mathrm{int}}(t)$ (typically charge/flux noise), the apparent length of the cavity fluctuates and destabilizes the resonant frequency from its desired point of operation $\omega_0(b_0)$, where we take $b_0$ to be the bias magnitude at the sample at time $t_0$. The cavity is driven using a carrier signal $\omega_\mathrm{c} = \omega_0(b_0)$ phase modulated with a modulation amplitude $\beta$ and modulation frequency $\omega_\mathrm{m}$ several times larger than the cavity linewidth $\kappa_{\mathrm{tot}}$. As we are particularly interested in cases where the cavity is driven at very low pump powers, we will follow the operator scattering approach used in Ref. \cite{kanhirathingal2021charge} to describe the resulting system dynamics. 

Treating the system semiclassically, the driving signal is described using $\langle a_p^{\mathrm{in}}(t) \rangle$, where $a_p^{\mathrm{in}}(t)$ is the annihilation operator of the transmission line input. Phase modulation of the carrier signal transforms the drive as below:
\begin{equation}
\begin{split}
    \langle a_p^{\mathrm{in}}(t) \rangle & =  \sqrt{\frac{P_p^{\mathrm{in}}}{\hbar \omega_\mathrm{c}}} e^{-i(\omega_c t+\theta_c)} \\ &
    \to \sqrt{\frac{P_p^{\mathrm{in}}}{\hbar \omega_\mathrm{c}}} \sum_{k=-\infty}^{k=\infty} J_k(\beta) e^{-i(\omega_c+k \omega_\mathrm{m})t},
    \end{split}
\end{equation}
where we have applied the Jacobi-Anger expansion to the exponential of the pump phase $\theta_{\mathrm{c}} = \beta \sin(\omega_\mathrm{m} t)$, with $P_{\mathrm{in}}$ the average pump power, and where $J_k$ is the Bessel function of the first kind. In Fig. \ref{basic_circuit_scheme}(a), we denote the input signal using its spectral components as $a_p^{\mathrm{in}}(\omega-\omega_{\mathrm{c}}, k \omega_{\mathrm{m}})$. We have adopted this notation everywhere in the figure to indicate that the signal is centered around the reference frequency described in the first argument. Thus for the case of $a_p^{\mathrm{in}}(\omega-\omega_{\mathrm{c}}, k \omega_{\mathrm{m}})$ the signal is centered around $\omega_{\mathrm{c}}$, and contains sidebands at the second argument $k\omega_\mathrm{m}$.

Since the sidebands lie outside the cavity linewidth, the phase of the delayed reflected signal at $\omega_{\mathrm{c}}$ interferes with these sideband signals after exiting the cavity. The steady-state system dynamics can be obtained from the quantum Langevin equation
\begin{equation}
   \dot{a}(t) =  -i\omega_0(t)a(t) - \frac{\kappa_{\mathrm{tot}}}{2}a(t)-i\sqrt{\kappa_{\mathrm{ext}}}a_p^{\mathrm{in}}(t),
\label{ql_eqn}
\end{equation}
where $a(t)$ is the cavity annihilation operator, and $\kappa_{\mathrm{tot}} = \kappa_{\mathrm{int}}+\kappa_{\mathrm{ext}}$ is the total damping rate, with $\kappa_{\mathrm{int}}$ and $\kappa_{\mathrm{ext}}$ the internal and external damping rates, respectively. Assuming $\delta b_{\mathrm{int}}(t) \ll b_0$, the fluctuating resonance $\omega_0(t) = \omega_\mathrm{c} + \delta \omega_0(t)$ takes the form
\begin{equation}
\label{omega_b_eqn}
    \omega_0(t) = \omega_\mathrm{c}+ g_b \delta b_{\mathrm{int}}(t),
\end{equation}
where we define $g_b$ as the coupling coefficient to the bias parameter $b$: $g_b = \left.\left(d\omega_0/db\right)\right|_{b=b_0}$. Using the transformation $\Tilde{a}(t) = a(t)\, e^{i\delta\omega_0t}$~\cite{wang2017quantum} corresponding to the rotating frame defined by the fluctuations $\delta b_{\mathrm{int}}(t)$, we can first modify the quantum Langevin equation in Eq. (\ref{ql_eqn}). By further applying the solution ansatz $\Tilde{a}(t) = \Tilde{\alpha}(t) \mathrm{exp} [-i\omega_\mathrm{c}t-\kappa_{\mathrm{tot}}t/2]$ into this equation,  we obtain for $\langle a(t) \rangle$:
\begin{equation}
    \langle a(t) \rangle =-i
    \sqrt{\frac{P_p^{\mathrm{in}}\kappa_{\mathrm{ext}}}{\hbar \omega_\mathrm{c}}} \sum_{k=-\infty}^{k=\infty} \frac{J_k(\beta) e^{-i(\omega_c+k \omega_\mathrm{m})t}}{i[\delta\omega_0(t)-k\omega_\mathrm{m}]+\kappa_{\mathrm{tot}}/2},
\end{equation}
where we neglect the contributions from the term containing $d\delta\omega_0/dt$ as $(d\delta\omega_0/dt) \Delta t \ll \delta\omega_0(t)$ in the ns time scale for $\Delta t$, compared to slowly changing fluctuations in resonance. Next, we obtain the output field $\langle a_p^{\mathrm{out}}(t) \rangle$ using the input-output relation $a_p^{\mathrm{out}}(t) = a_p^{\mathrm{in}}(t) - \sqrt{\kappa_{\mathrm{ext}}} a(t)$~\cite{gardiner-input-1985}:
\begin{equation}
    \langle a_p^{\mathrm{out}}(t) \rangle =
    \sqrt{\frac{P_p^{\mathrm{in}}}{\hbar \omega_\mathrm{c}}} \sum_{k=-\infty}^{k=\infty} r_k(t) J_k(\beta) e^{-i(\omega_c+k \omega_\mathrm{m})t},
\end{equation}
where $r_k(t)$ can be written as,
\begin{equation}
\label{rn_eq}
    r_k(t) = \frac{k\omega_\mathrm{m} - \delta \omega_0(t) + i(\kappa_{\text{int}}- \kappa_{\text{ext}})/2}{k\omega_\mathrm{m} - \delta \omega_0(t) + i(\kappa_{\text{int}}+ \kappa_{\text{ext}})/2}.
\end{equation}
Notice that $r_k(t)$ takes the general form of a reflection-coefficient at $\omega_c+k\omega_\mathrm{m}$, but is slowly time-varying due to the low-frequency fluctuations in cavity resonance itself. Typical measurements of cavity reflection coefficients using a vector network analyzer outputs the value averaged over the measurement time, and often smears out the effects of these resonant fluctuations~\cite{brockfrequency2020,neill2013fluctuations}.

The output power can be obtained using $\langle P_p^{\mathrm{out}}(t)\rangle =\langle V_p^{\mathrm{out}}(t)\rangle^2/Z_p$, where $Z_p$ is the transmission line impedance and $V_p^{\mathrm{out}}(t)$ is the output voltage given by the following:~\cite{kanhirathingal2021charge}
\begin{equation}
    \begin{split}
        V_p^{\mathrm{out}}(t) =-i \int_0^{\infty} d\omega  \sqrt{\frac{\hbar \omega}{4 \pi Z_p}}\; & \left[ e^{-i\omega t} a_p^{\text{out}}(\omega)-\right.\\ &\left. e^{i\omega t} \left(a_p^{\text{out}}(\omega)\right)^{\dagger}\right].
    \end{split}
\end{equation}
Note that the output power spectral components have an implicit dependence on time due to the low-frequency fluctuations $\delta b_{\mathrm{int}}(t)$ of the bias parameter. Furthermore, detection of $\langle P_p^{\mathrm{out}}(t)\rangle$ will result in an oscillating signal with frequencies $k\omega_\mathrm{m}$. The DC component of this signal gives the reflected intensity, and has its minimum at $\delta \omega_0 = 0$, with a symmetric response about this point. We are, however, interested in measuring the contribution oscillating at $\omega_\mathrm{m}$, which can be obtained as
\begin{equation}
\begin{split}
\label{Pout_eq}
    \mathcal{P}(t) = & J_0(\beta) J_1(\beta) P_p^{\mathrm{in}} \\ & \left[e^{i\omega_\mathrm{m} t} \left(r_0(t)r^*_1(t)-r^*_0(t)r_{-1}(t)\right)+\right. \\ & \left. e^{-i\omega_\mathrm{m} t} \left(r^*_0(t)r_1(t)-r_0(t)r^*_{-1}(t)\right)\right],
    \end{split}
\end{equation}
where we have neglected the contribution from the second harmonics and above assuming smallness of $J_k(\beta)$ for $k>1$.
\begin{figure*}[thb]
\centering
\includegraphics[totalheight=0.2\textheight, trim = 200 0 70 0]{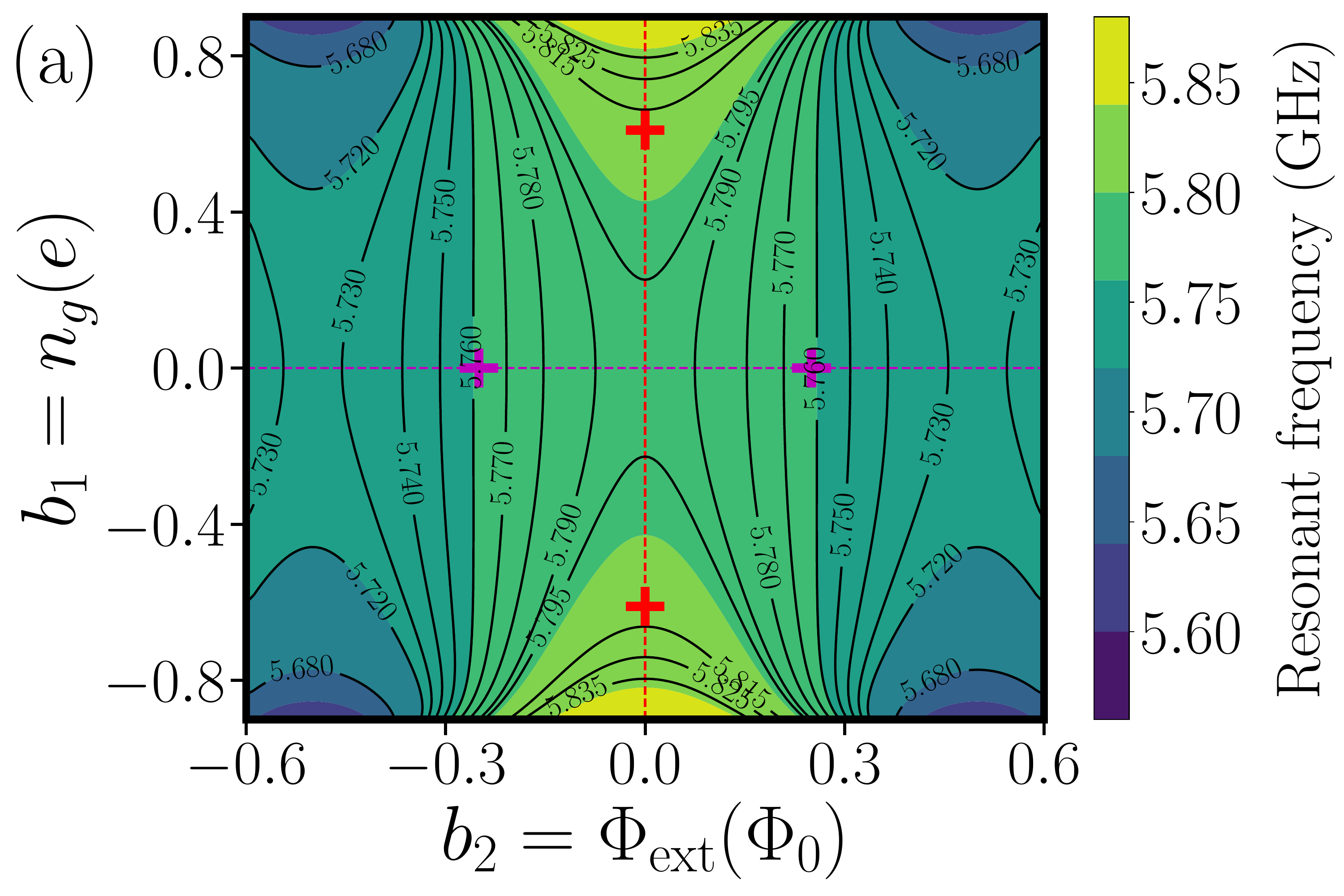}
\includegraphics[totalheight=0.2\textheight]{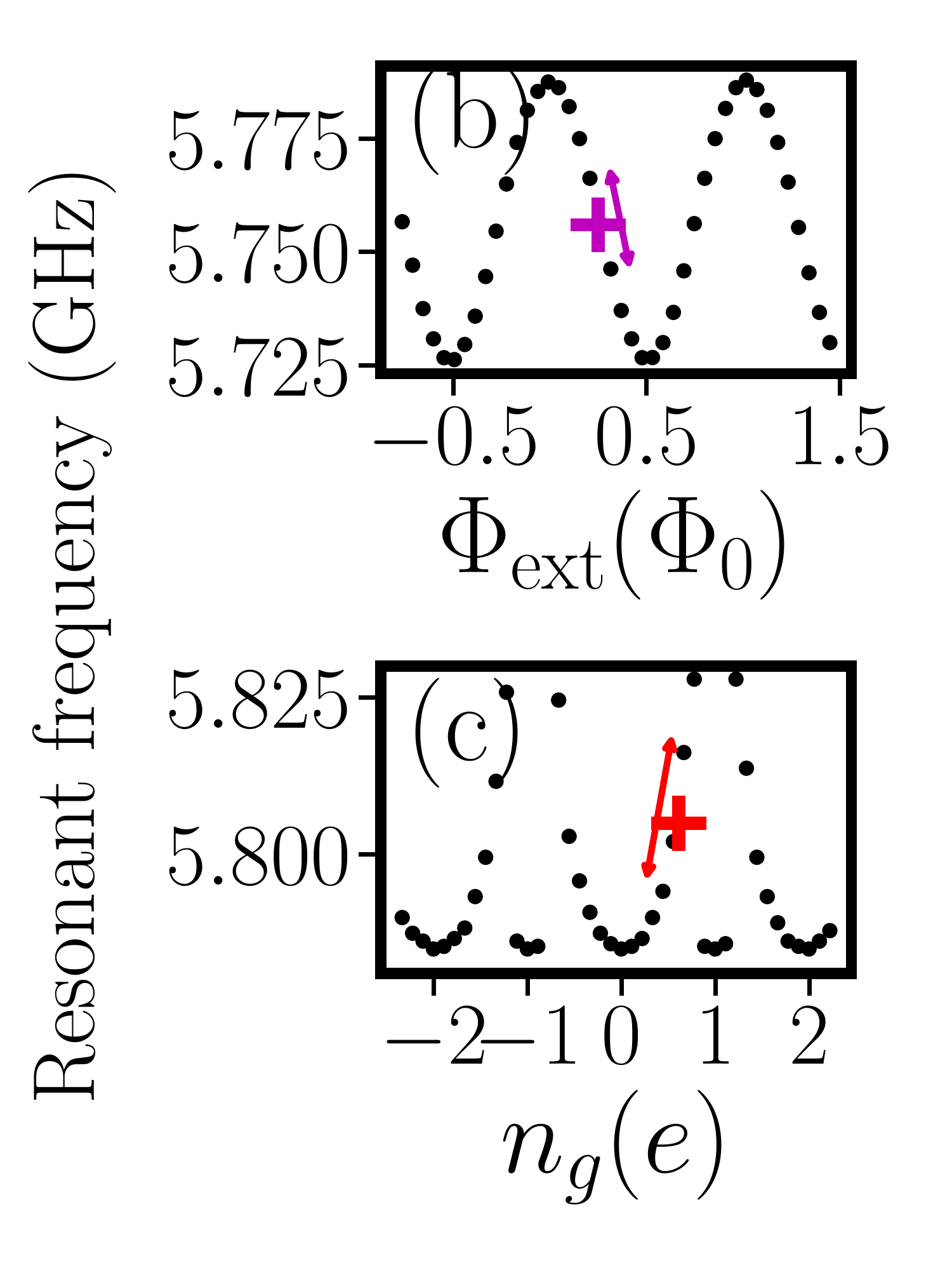}
\includegraphics[totalheight=0.2\textheight, trim = 0 0 200 0]{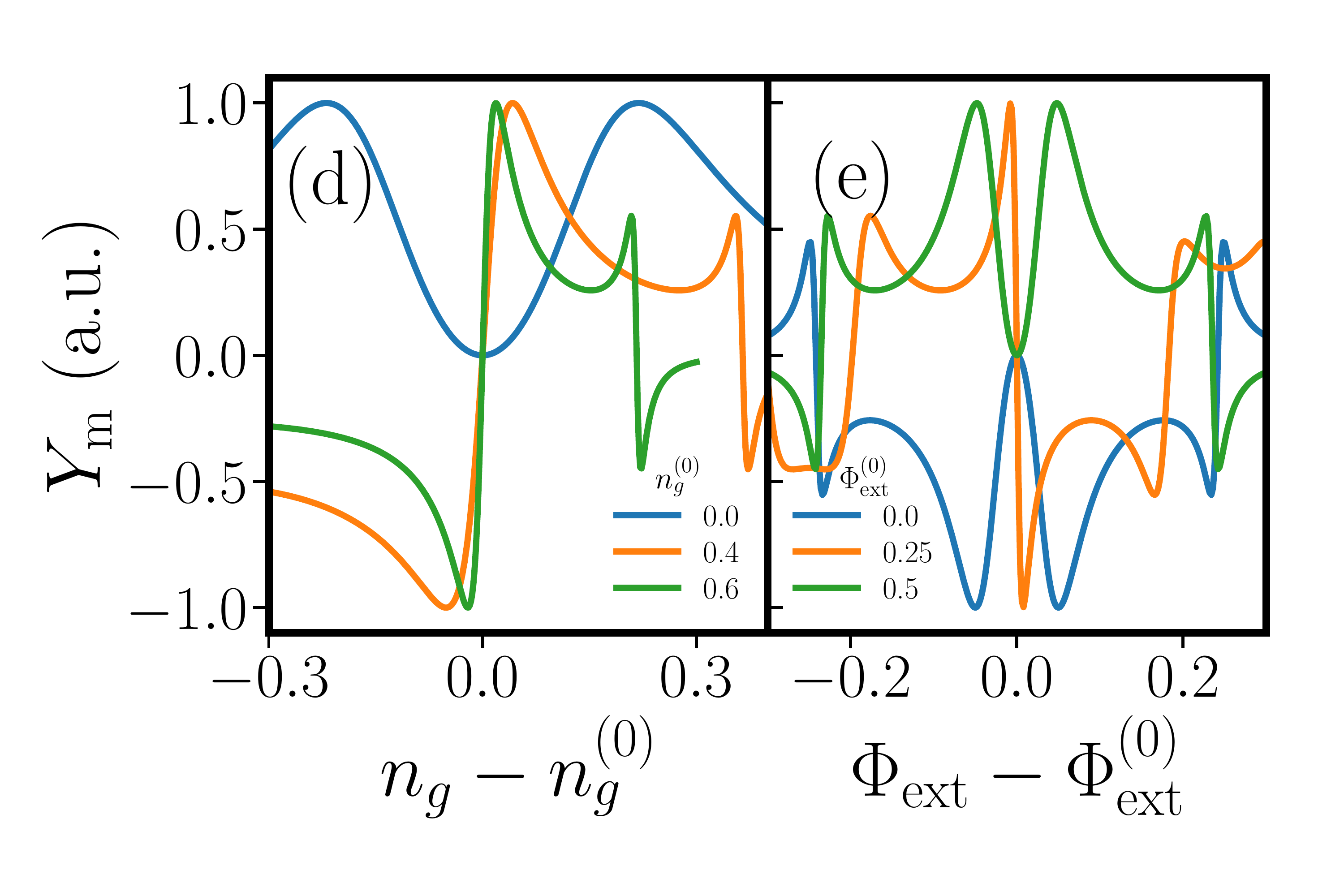}
\caption{(a) Contour plot displaying the resonant frequency $\omega_0$ as a function of the two-dimensional bias space. We avoid feedback locking in regions where the cavity is sensitive to both bias parameters simultaneously, so as to avoid accidentally destabilizing the cavity away from the bias point of interest. (b) Measured resonant response as a function of flux along $n_g=0$ (purple line in (a)), where the charge noise is minimal. The purple plus sign denotes point of maximum flux sensitivity. (c) Measured resonant response as a function of gate along $\Phi_{\mathrm{ext}}=0$, where the flux sensitivity is minimal (red line in (a)). The charge sensitivity increases towards charge degeneracy ($n_g=1$), but we avoid operating the feedback loop in the region $|n_g|>0.65$ because of quasiparticle poisoning. The red plus sign denotes point of high gate sensitivity. (d) Simulated $Y_\mathrm{m}$($n_g$) calculated about different bias values of $n_g^{(0)}$. The monotonicity is steeper for higher values of $n_g$ and non-existent at $n_g=0$. (e) Simulated $Y_\mathrm{m}$($\Phi_{\mathrm{ext}}$) calculated about different bias values of $\Phi_{\mathrm{ext}}^{(0)}$. Unsuitable points of feedback operation are near $\Phi_{\mathrm{ext}}^{(0)} = 0$ and $\Phi_{\mathrm{ext}}^{(0)}=0.5\Phi_0$.
\label{fig:fig_ccpt_sim}}
\end{figure*}
The above expression represents the first derivative of the reflected intensity encoded in the first sideband $\omega_{\mathrm{m}}$, which will behave as a monotonically increasing function with its zero at $\delta\omega_0=0$. Thus, as long as ($d\omega_0/db$) varies monotonically as a function of $b$, we can utilize this information as a potential error signal to counteract the fluctuations $\delta b_{\mathrm{int}}(t)$. In Fig. \ref{basic_circuit_scheme}(a), the output signal is denoted by $a_p^{\mathrm{out}}(\omega-\omega_{\mathrm{c}}, k \omega_{\mathrm{m}}, \delta b_{\mathrm{int}})$, indicating that the signal is centered about $\omega_{\mathrm{c}}$, has sidebands at $k\omega_\mathrm{m}$, and contains information about the intrinsic bias noise. Similarly, the $\mathcal{P}(\omega-k\omega_\mathrm{m},\delta b_{\mathrm{int}})$ term in Fig. \ref{basic_circuit_scheme}(a) indicates that the signal has frequencies at $k\omega_\mathrm{m}$.

To maximize the sensitivity in detecting the fluctuations $\delta \omega_0(t)$, we utilize the sine quadrature of expression (\ref{Pout_eq}). We achieve this experimentally using a lock-in amplifier with the reference signal taken from the original modulation source. Quantitatively, the in-phase cosine quadrature $X_\mathrm{m}(t)$ is insensitive to these resonant fluctuations while the sine quadrature $Y_\mathrm{m}(t)$ is given by~\cite{burnett2013high}
\begin{equation}
\begin{split}
\label{yquadeqn}
        Y_\mathrm{m}(t) = \;& 2 J_0(\beta) J_1(\beta) P_p^{\mathrm{in}} \\ & \Big(
        \operatorname{Im} [r_0(t)]
         \big( \operatorname{Re} [r_1(t)]+ \operatorname{Re} [r_{-1}(t)]\big) - 
         \\
         & \operatorname{Re}[r_0(t)] 
          \big(\operatorname{Im} [r_1(t)]+ \operatorname{Im}[r_{-1}(t)]\big)
          \Big).
\end{split}
\end{equation}
For $\omega_{\mathrm{m}}$ well outside the cavity line-width, $\operatorname{Im}[r_{\pm 1}(t)] \to 0$ and $\operatorname{Re}[r_{\pm1}] \to 1$ for cavity resonances $\omega_0(b)$ in the vicinity of $\omega_{\mathrm{c}}$. In short,
\begin{equation}
    Y_\mathrm{m}(t) = 16 \kappa_{\mathrm{ext}} J_0(\beta) J_1(\beta) P_p^{\mathrm{in}} \; \frac{g_b \delta b_{\mathrm{int}}(t)}{\kappa_{\mathrm{tot}}^2 + 4g_b^2\delta b_{\mathrm{int}}(t)^2}.
\end{equation}
We obtain the bandwidth of the monotonic region of $Y_\mathrm{m}(t)$ to be $\kappa_{\mathrm{tot}}$ using the condition
$d\operatorname{Im}[r_0]/d\delta\omega_0 = 0.$

The normalized transfer function of the lock-in amplifier is that of a single-pole, low-pass filter given by $G_{\mathrm{LA}}(\omega) = (1+i\omega\tau_{\mathrm{LA}})^{-1}$, where $\tau_{\mathrm{LA}}$ is the lock-in amplifier time-constant. Hence in the region where the approximation $\delta b_{\mathrm{int}} \ll \kappa_{\mathrm{tot}}/2g_b$ is valid, the net transfer function of the open-loop setup can be written as
\begin{equation}
    G(\omega) = \frac{Y_\mathrm{m}(\omega)}{\delta b_{\mathrm{int}}(\omega)} = \frac{G_0 }{1+i\omega\tau_{\mathrm{LA}}},
\end{equation}
where $G_0$ is the net gain,
\begin{equation}
\label{G0eqn}
    G_0 = \left(\frac{4J_1(\beta)}{J_0(\beta)}\right) n \hbar \omega_0(b_0) g_b G_{\mathrm{amp}},
\end{equation}
and we have expressed the input power in terms of $n = 4 \kappa_{\mathrm{ext}} J_0^2(\beta) P_p^{\mathrm{in}}/\hbar \omega_0(b_0) \kappa_{\mathrm{tot}}^2$, the average number of photons in the cavity. Here, $G_{\mathrm{amp}}$ is the net gain of the amplifier chain, including that of the power detector and the lock-in amplifier. Note that we have neglected the secondary fluctuations of $n$ and $\omega_0$ in the above expression, induced due to the bias fluctuations.

Moreover, the PSD of fluctuations in $Y_\mathrm{m}(t)$ takes the form
\begin{equation}
\label{syy_eqn}
    S_{Y_\mathrm{m}Y_\mathrm{m}}(\omega) = \left(\frac{G_0^2}{1+\omega^2\tau_{\mathrm{LA}}^2}\right) \; S_{bb}(\omega),
\end{equation}
where $S_{bb}(\omega)$ is the PSD of the bias noise.

We can now close the feedback loop in our setup by applying a control law $K(\omega)$ through a PID controller, such that $Y_\mathrm{m}(t)$ follows the control signal $Y_{\mathrm{ref}} = Y_\mathrm{m}(t_0)$ as closely as possible. For cavities with linear reflection coefficients discussed above, this control value is zero, as can be determined from Eq. (\ref{yquadeqn}). The closed-loop transfer function takes the form $K(\omega)G(\omega)/\left(1+K(\omega)G(\omega)\right)$. Using the condition $K(\omega)G(\omega) \gg 1$ and $K(\omega)$ chosen such that the loop does not pick up substantial sensor noise~\cite{bechhoefer2005feedback}, we can thus compensate for the fluctuations $\delta b_{\mathrm{int}}(\omega)$ up to a bandwidth of $1/\tau_{\mathrm{LA}}$.
\section{Application to the \lowercase{c}CPT}
\label{ccpt_sec}
The scenario discussed in the previous section is frequently observed in many open quantum systems, where the tunability control of the system of interest introduces noise and results in reduced measurement sensitivity or in some cases, decreased coherence properties~\cite{burnett2019decoherence,gyenis2021experimental, bylander2011noise}. In this section we discuss the implementation of the scheme presented in Sec. \ref{concept} to one such system, the cCPT [presented in Figs. \ref{basic_circuit_scheme}(b) and (c)]. Similar to the system described in Fig. \ref{basic_circuit_scheme}(a), the cCPT device communicates with the external pump/probe setup through its quarter-wave superconducting microwave resonator. The non-linear Josephson inductance emerging from the Cooper pair transistor introduces two-dimensional tunability to the resonance, either via the gate voltage $V_g$ controlling the island charge of the Cooper pair transistor, or via the external flux bias $\Phi_{\mathrm{ext}}$, coupling the cavity phase and the differential phase of the Josephson junctions via a SQUID loop; the current work mainly focuses on the suppression of the resonant frequency fluctuations caused due to charge noise coupling to the cavity at low frequencies. The resulting reduction of the $1/f$-noise, as detailed in Sec. \ref{results}, is significant enough to potentially elevate the cCPT to operate in an ultra-sensitive regime for electrometry.

Following the formalism in Sec. \ref{concept}, we now have the bias vector $\vec{b} = (n_g, \Phi_{\mathrm{ext}})$ and the resonant frequency shift $\delta \omega_0(\vec{b})$ inversely proportional to the Josephson inductance $L_{\mathrm{CPT}}$ given by~\cite{kanhirathingal2021charge,brock2021nonlinear}
\begin{eqnarray}
    L_{\mathrm{CPT}}^{-1} = \frac{\partial^2 E_{\mathrm{CPT}}^{(0)}}{\partial b_2^2},
    \label{CPTLeq_main}
\end{eqnarray}
where $E_{\mathrm{CPT}}^{(0)}$ is the ground state energy of the CPT described by the Hamiltonian with matrix coefficients 
\begin{equation}
    \langle N|H_{\mathrm{CPT}}|N\rangle = 4E_c \left(N-\frac{b_1}{2}\right)^2,
\end{equation}
and
\begin{equation}
    \langle N|H_{\mathrm{CPT}}|N+1\rangle = \langle N|H_{\mathrm{CPT}}|N-1\rangle = E_J(b_2),
\end{equation}
where $E_c$ and $E_J(b_2=\Phi_{\mathrm{ext}})$ are the charging and the Josephson energies of the CPT, respectively. The ket $|N\rangle$ denotes the number of excess Cooper pairs on the CPT island and the gate polarization number $b_1=n_g$ is related to the externally applied gate voltage $V_g$ via $n_g = C_g V_g/e$.

Fig. \ref{fig:fig_ccpt_sim}(a) provides a simulated 2-D plot of the tunable resonant frequency based on the experimental characterization of the cCPT. As can be seen in this contour plot, a single value of $\omega_0$ can correspond to a continuum of possible values in the bias space. The feedback scheme corrects for the bias fluctuations purely based on the detuning of the carrier signal from the resonance. As a result, applying the technique to a simultaneous charge and flux sensitive region can result in increased instability in the applied bias along a contour while still stabilizing the resonant frequency fluctuations. We therefore limit our measurements (presented in later sections) in the regimes where the cCPT is sensitive to one of the bias parameters while minimizing the coupling to the other ones.

Figs. \ref{fig:fig_ccpt_sim}(b) and \ref{fig:fig_ccpt_sim}(c) provide the measured frequency response around these bias-sensitive regimes. Fig. \ref{fig:fig_ccpt_sim}(b) plots $\omega_0(b_2)$ while $b_1=0$ such that the gate is effectively decoupled from the cavity. Similarly, Fig. \ref{fig:fig_ccpt_sim}(c) plots $\omega_0(b_1)$ while $b_2$ is set to zero, i.e., with minimal flux noise. Notice that for $0.1 \leq |n_g| \leq 0.65$, $n_g$ corresponds to frequency shifts that are monotonic on the order of tens of MHz - several times larger than the typical cavity linewidths. Thus our feedback scheme can be applied across an appreciable span along $n_g$. The region $|n_g|>0.65$ is highly prone to quasi-particle poisoning, and we avoid operations in this regime, as discussed later in Sec. \ref{results}. The simulated $Y_\mathrm{m}(b)\equiv Y_\mathrm{m}[\omega_0(b)]$ responses are plotted in Figs. \ref{fig:fig_ccpt_sim}(d) and \ref{fig:fig_ccpt_sim}(e), which captures the effects of the shift of the applied bias $b$ from $b_0$. As expected, for the case of $n_g^{(0)} = 0$, $Y_\mathrm{m}(n_g)$ is symmetric about $n_g-n_g^{(0)}=0$, i.e., it does not have a one-to-one mapping onto its respective bias value making this the regime unsuitable for the feedback application. Similar conclusions about feedback applicability in flux noise suppression can be deduced from Fig. \ref{fig:fig_ccpt_sim}(e).

The results reported in this work also involve driving the cavity to Kerr-shifted regimes. The resulting non-linear reflection coefficient takes the form of Eq. (\ref{rn_eq}) with $\delta \omega_0 \to \delta \omega_0+Kn(\delta \omega_0)$, where $K$ is the Kerr-coefficient and $n(\delta \omega_0)$ is the number of average photons in the cavity given by the roots of the following equation:~\cite{brock2021nonlinear}
\begin{equation}
    n^3K^2 + 2\delta \omega_0 K n^2 + [\delta \omega_0^2 + \kappa_{\mathrm{tot}}^2/4]n - \kappa_{\mathrm{ext}} P_{\mathrm{in}}/\hbar\omega_0 = 0.\\
\end{equation}

As the Kerr-coefficient can be strong enough to produce a Kerr-shift comparable to the cavity linewidth of the cCPT for $n \geq 5$, it is important to look at its effects on the error signal generation. Depending on the specific application of interest, we may require driving the cavity  exactly at linear resonance with $\omega_{\mathrm{c}} = \omega_0(b_0)$. The reference signal $Y_{\mathrm{ref}}(b_0)$ in this case corresponds to a non-zero value, and the Kerr-induced asymmetries in $r_n(t)$ can be strong enough for the error signal to not behave monotonically about the resonance. The former merely requires a recalibration at each bias point, while the latter effectively acts as an upper bound in limiting the application of the feedback technique at higher input powers. But the feedback scheme can still be applied at these input powers with $Y_{\mathrm{ref}} = 0$, provided the carrier signal is set to the point of minimum reflection coefficient, given by $\omega_\mathrm{c} = \omega_0(b_0)+nK$.
\section{Experimental Setup}
\label{experiment}
\begin{figure}[thb]
\includegraphics[width = 0.45 \textwidth, trim = 0 0 0 0]{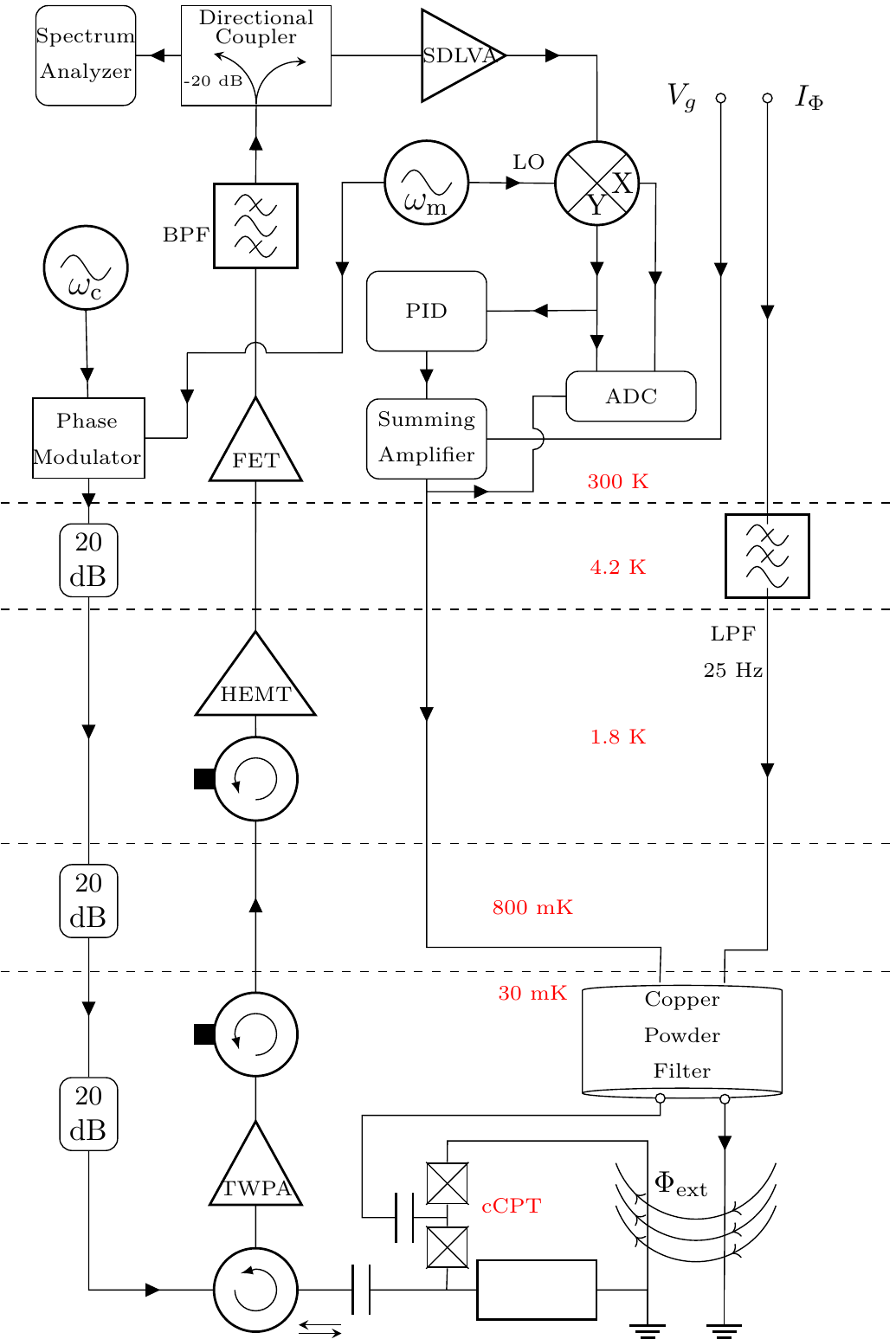}
\caption{\label{circuit_diagram}Experimental setup for the dynamic feedback control of the intrinsic bias noise coupling to the cavity.}
\end{figure}

We present in this section the experimental realization of the scheme discussed in the previous sections. The underlying circuitry behind the detection of the error signal is similar to the Pound-Drever-Hall technique applied to superconducting microwave resonators~\cite{lindstrom2011pound}. In contrast to the conventional technique, which corrects the drive frequency, we use the PID output to change the bias parameter, thereby stabilizing the resonant frequency of the cavity itself. The circuitry enabling such a measurement is shown in Fig. \ref{circuit_diagram}, and is detailed in the following.
\subsection{Circuitry}
The input drive consists of a carrier signal $\omega_{\mathrm{c}}$ at the cavity resonance frequency, which is phase-modulated (using an Analog Devices HMC-C010 phase-shifter) at a frequency $\omega_{\mathrm{m}}$. The reflected output signal is amplified at different stages and is sent into a directional coupler where the signal is to split into two routes: the feedback loop component A and the actual measurement component B. The -20 dB coupled port sends signal B to a spectrum analyzer, which can be used to track the power spectral components when the feedback loop is active. Signal A enters a highly sensitive power detector (SDLVA HMC-C088), which outputs a voltage proportional to the input power, with frequency components at the harmonics of $\omega_\mathrm{m}$. The lock-in amplifier then mixes this signal with the reference signal at $\omega_\mathrm{m}$ to output the two quadrature components. The error signal of interest is contained in the Y-quadrature such that a fluctuation of the cavity resonance frequency is typically measured as a non-zero value [see Fig. \ref{fig:fig_ccpt_exp}(b)]. When the cCPT is biased at points where flux/charge causes the dominant source of intrinsic noise, we attribute these measured resonance frequency fluctuations to disturbances in that bias parameter. The output of the PID controller then corrects for the error by application to the bias parameter, which in our case is the gate voltage, via a summing amplifier. The summing amplifier is bandwidth-limited to 1 MHz. This reduces high-frequency noise, while allowing modulations for charge-sensitivity measurements up to a few 100 kHz.

The cCPT sample used for the following measurements exhibits a total tunability of about 140 MHz, centered about the bare cavity frequency at 5.757 GHz. Following a model that accounts for frequency fluctuations in the cavity~\cite{brockfrequency2020}, the typical external and internal damping rates observed at $(n_g, \Phi_{\mathrm{ext}}) = (0,0)$ are $\sim$0.97 MHz and $\sim$0.3 MHz, respectively. We therefore fix the modulation frequency $\omega_\mathrm{m}$ to be 30 MHz, one order of magnitude higher than the total damping rate. The average photon numbers reported in this paper are calculated employing a model that considers the linear relation between the input power at the sample and the associated Kerr-shift in the cavity resonance frequency~\cite{brock2021nonlinear}.
\subsection{Benchmarking }
\begin{figure*}[thb]
\hspace{-1cm}
\includegraphics[height=0.2\textheight, trim = 0 0 0 0]{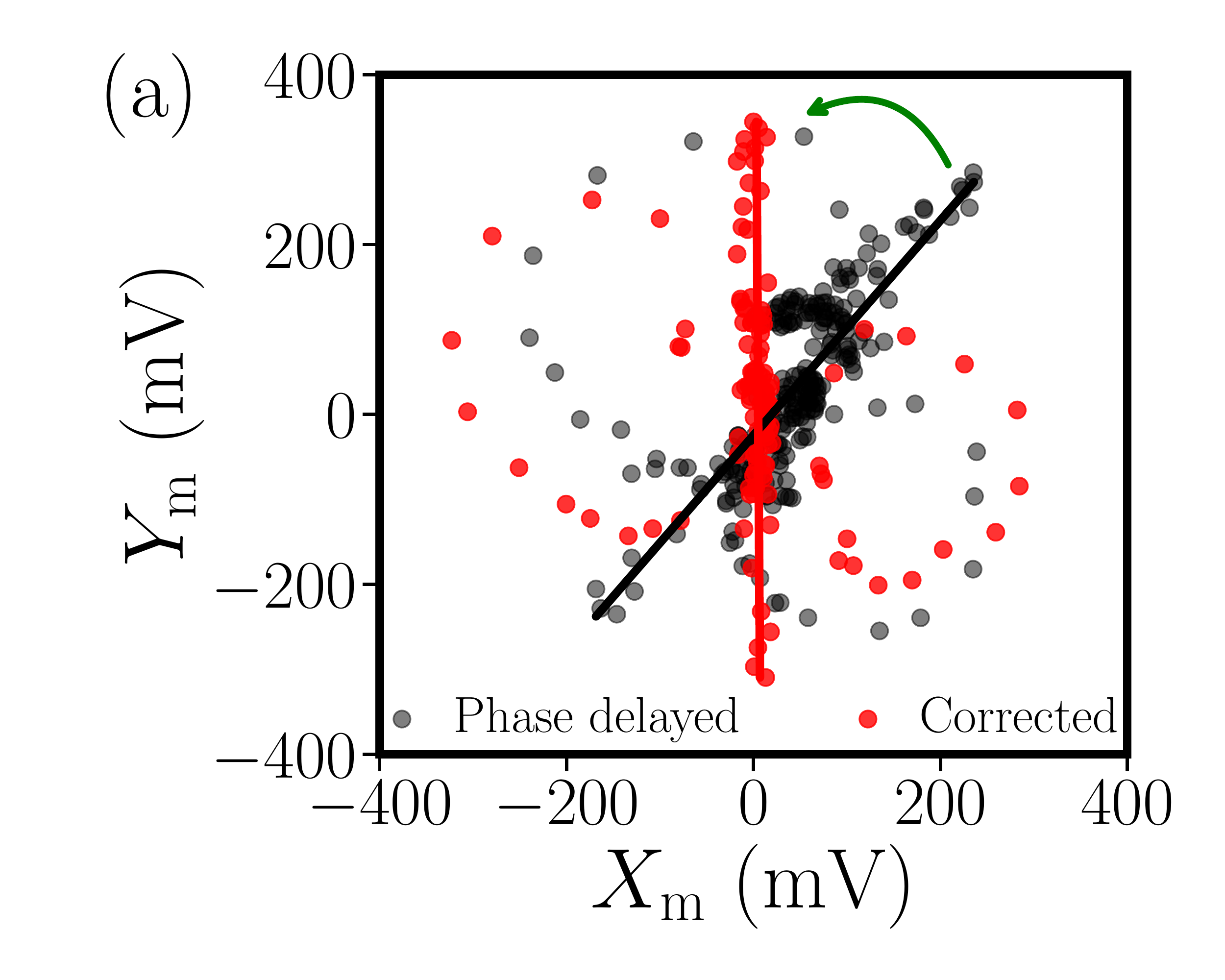}
\includegraphics[height=0.2\textheight, trim = 0 0 0 0]{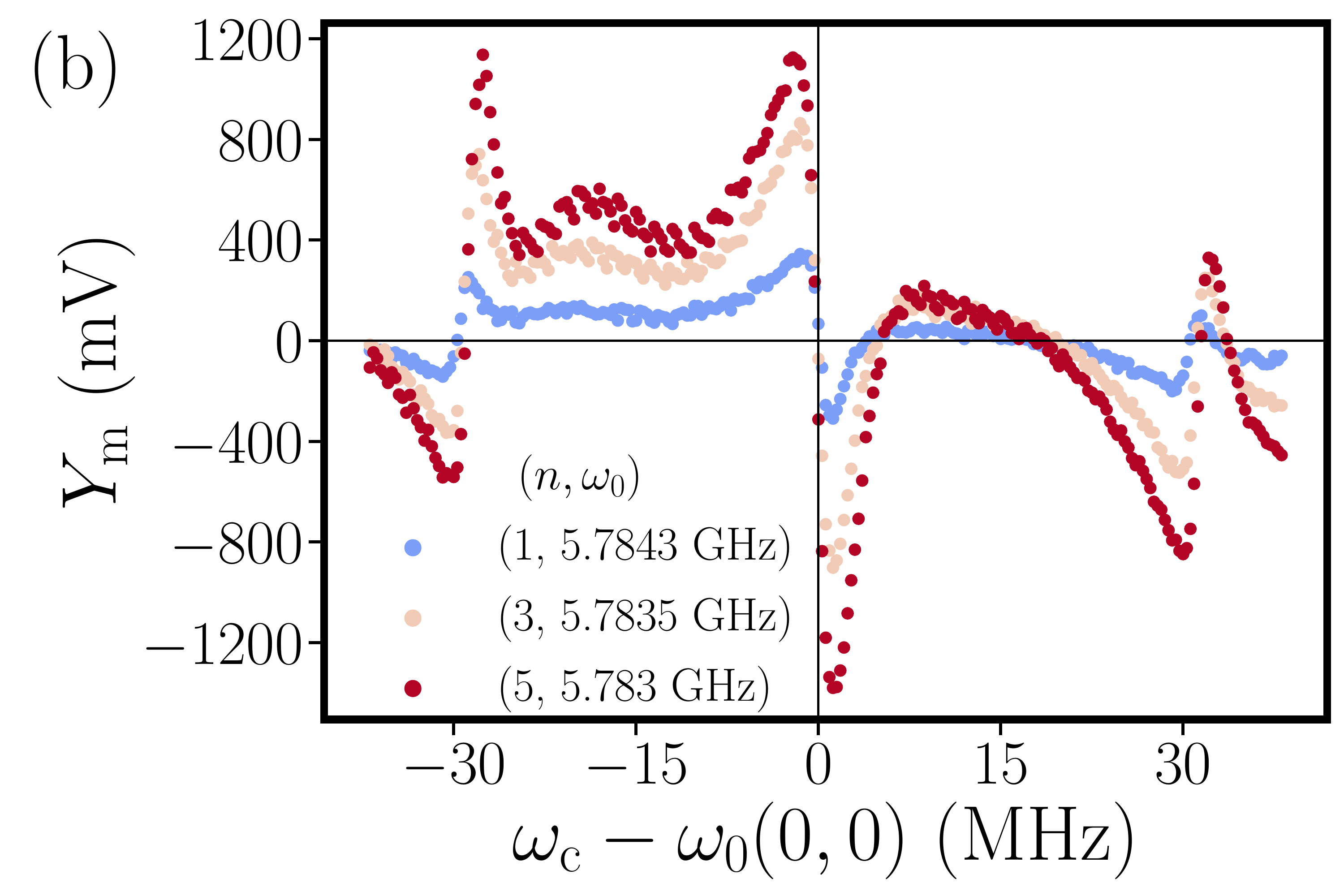}\\
\vspace{0.2cm}
\includegraphics[height=0.2\textheight, trim = 20 0 0 0]{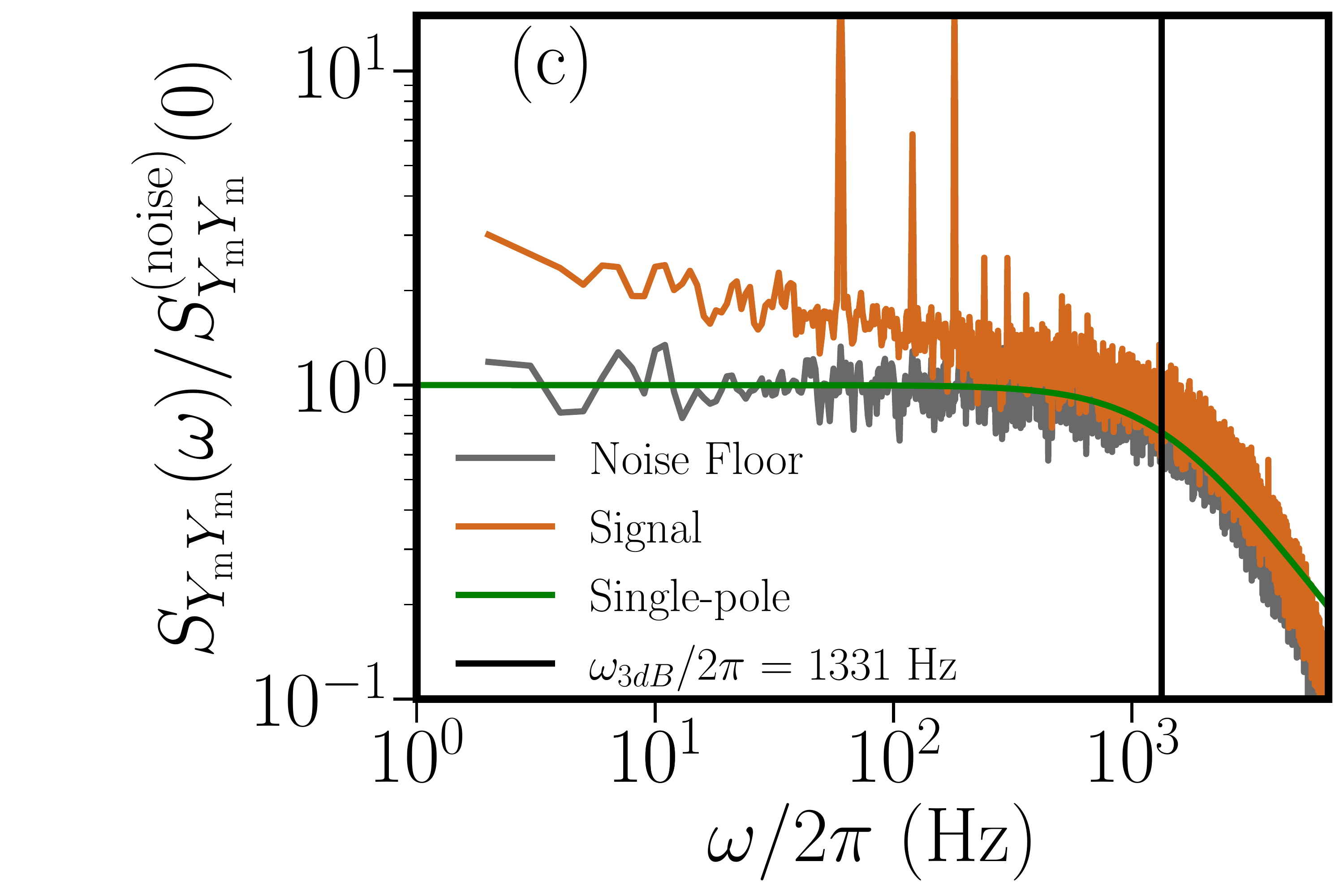}
\includegraphics[height=0.225\textheight, trim = 0 0 0 0]{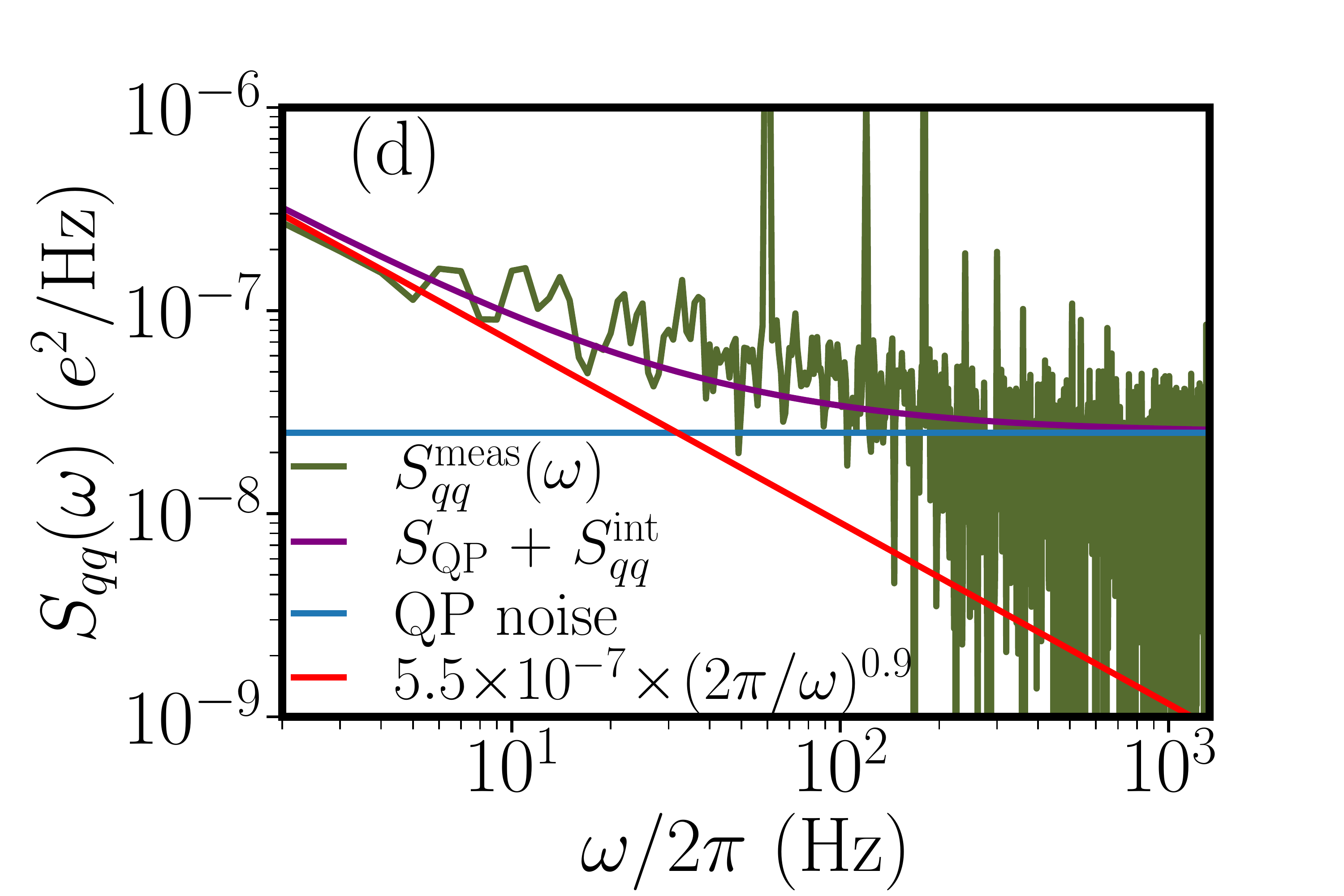}
\caption{(a) Quadrature outputs from the lock-in amplifier displayed as a parametric plot. The data is measured for varying $\omega_\mathrm{c}$, with resonance fixed at $\omega_0$(0,0), where both the flux and charge noise is a minimum for the cCPT, and the cavity is driven at photon number $n=1$. The black plot represents the data before the phase delay correction. All the feedback measurements are carried out with the phase of the reference signal set to the one in the red plot. This ensures that the error signal has maximum sensitivity to fluctuations. (b) Measured $Y_\mathrm{m}(\omega_c)$ as a function of the detuning $\omega_\mathrm{c}$-\;$\omega_0(0,0)$ and varying photon number $n$, where $\omega_0(0,0)$ is the Kerr-shifted resonance. The zero-point of the error signal corresponds to the Kerr-shifted resonance value, enabling us to set the reference value for feedback as zero, even in strongly Kerr-nonlinear regimes. Each point is the average of a 1 sec acquisition with sampling rate of 1 kHz. The time constant of the lock-in amplifier is set to 10 ms to average out fluctuations and improve resolution. The data in (a) and (b) are taken for $\beta=1.08$. (c) The PSD of fluctuations in $Y_\mathrm{m}(t)$ given by $S_{Y_\mathrm{m}Y_\mathrm{m}}(\omega)$ for the open-loop setup. The grey plot displays the noise floor measured at the lock-in amplifier sine quadrature output. The green plot is the single-pole, low-pass filter fit applied to the above data. The cut-off frequency obtained is 1331 Hz, set by the lock-in amplifier time constant. The orange plot captures the charge fluctuations when the cCPT is biased in the increased charge-sensitive regime $(n_g,\Phi_{\mathrm{ext}})$ = (0.6, 0), and the cavity is driven at $n=10$. The measurement is completed in 10 sec with a sampling rate of 100 kHz. The data displayed in the plot is scaled to the amplitude of the noise floor to better indicate the SNR. (d) PSD of the charge noise calculated for the data in (c). $S^{\mathrm{meas}}_{qq} (\omega)$ is the total charge noise with contributions from the intrinsic charge noise fluctuations $S_{qq}^{\mathrm{int}}$ at the CPT (red plot varying as $\sim1/f$), and the fluctuations $S_{\mathrm{QP}}$ due to quasiparticle switching with a Lorentzian noise floor (blue plot). Note the Lorentzian floor appears as white noise as the roll-off frequency for Lorentzian fit is not resolvable using this measurement. The purple plot corresponds to the net fit $S_{\mathrm{QP}}$+$S_{qq}^{\mathrm{int}}$.}
\label{fig:fig_ccpt_exp}
\end{figure*}

Since the measurements are performed in the few-photon limit, we have optimized our setup at each stage to attain the maximum signal-to-noise ratio (SNR) at the output. Firstly, as the magnitude of the error signal is proportional to $J_1(\beta)/J_0(\beta)$ [refer Eq. (\ref{G0eqn})] for a fixed average photon number in the cavity, we choose $\beta = 1.84$ to provide increased sensitivity. This value is chosen such that $J_1(\beta)$ is maximized, and $J_0(\beta)$ is not too low a coefficient to achieve cavity driving.

The circuitry is further refined to ensure the error signal behaves in a manner discussed in Secs. \ref{concept} and \ref{ccpt_sec}. For example, Eq. (\ref{yquadeqn}) can also have contributions from the cross-terms involving sidebands at $\pm \omega_\mathrm{m}$ and $\pm 2\omega_\mathrm{m}$. A tunable bandpass filter with center frequency near resonance and bandwidth less than $4 \omega_\mathrm{m}$ is inserted after the room temperature amplifiers to partially filter out the DC and $2\omega_\mathrm{m}$ components. This prevents the saturation of the power detector and ensures a larger SNR at the power detector output by reducing the input noise~\cite{burnett2013high}. 

A near quantum-limited traveling wave parametric amplifier (TWPA)~\cite{macklin2015near} at the first-stage amplification improves the real-time detection of resonance frequency fluctuations at the single-photon level. For the efficient detection of the phase-modulated signal by the power detector, the bias power and frequency of the TWPA pump are chosen such that the mean SNR across the cCPT's tunable range is maximum, corresponding to a noise bandwidth of 80 MHz (equal to that of the tunable bandpass filter), and a signal of one photon. The gain profile also displays minimal ripples at these bias values to achieve relatively symmetric response at either of the sidebands. This ensures the error signal response is not influenced by the gain profile features, and the cavity response is closely tracked.

Since the output signal reflected from the cavity goes through several meters of cable and other microwave components as compared to the reference signal used by lock-in amplifier, the sine quadrature output is typically phase-shifted to a different quadrature. Hence, we correct for this phase delay using a frequency sweep of the carrier signal and simultaneous measurement of both quadratures, with the cCPT biased at the minimally flux and gate sensitive point $(n_g,\Phi_{\mathrm{ext}}) = (0,0)$. As shown in Fig. \ref{fig:fig_ccpt_exp}(a), a phase delay causes a rotation in the phase space, and can be corrected for accordingly. Fig. \ref{fig:fig_ccpt_exp}(b) constitutes an accurate representation of the sine quadrature as a function of carrier signal around resonance, after accounting for this correction, and for varying average photon number in the cavity. As can be seen, the zero-point of the error signal remains at the Kerr-shifted resonance value, allowing us to set the reference value for the feedback signal at zero, even when the cavity is driven into the Kerr-regime.

The fluctuations in $Y_\mathrm{m}(t)$ as measured by the digitizer, given by $S_{Y_\mathrm{m}Y_\mathrm{m}}(\omega)$, for the open-loop setup when the cavity is driven at $n=10$ is provided in Fig. \ref{fig:fig_ccpt_exp}(c). The PSD of the time-domain data collected over 10 sec at 100 kHz sampling rate is plotted in this figure. The data is scaled to the amplitude of the noise floor to clearly display the signal-to-noise ratio (SNR) of the measurement. The off-resonance noise measurement of the Y-quadrature of the lock-in amplifier outputs a single-pole, low-pass filter transfer function given by $G(\omega) = (1+i\omega/\omega_{\mathrm{LPF}})^{-1}$, where $\omega_{\mathrm{LPF}}=2\pi \times$ 1331 Hz,  close to the lock-in amplifier bandwidth set by the time constant 100 $\mu$sec. The time constant is set to measure a reasonable bandwidth of low-frequency fluctuations. A higher bandwidth detects more fluctuations but it necessitates an associated decrease in the measurement time, negatively affecting the SNR simultaneously. 

In order to calculate the PSD of the intrinsic charge noise $S_{qq}^{\mathrm{int}}(\omega)$, we first obtain the DC gain $G_0 = \left.G(\omega)\right|_{\omega=0}$. This is calculated from the slope of $\bar{Y}_\mathrm{m}(|\delta n_g| \leq 0.01)$, where $\bar{Y}_\mathrm{m}(|\delta n_g|)$ corresponds to the time-averaged value of $Y_\mathrm{m}(|\delta n_g|)$ in the vicinity of our bias point of interest, which for the case discussed in Fig. \ref{fig:fig_ccpt_exp}(d) is at $n_g = 0.6$. After accounting for the noise floor, we may utilize Eq. (\ref{syy_eqn}) to obtain the measured charge noise $S^{\mathrm{meas}}_{qq} (\omega)$. 

As described in Fig. \ref{fig:fig_ccpt_sim}(c), the cCPT is susceptible to quasiparticle poisoning (QP) for $n_g$ closer to charge degeneracy. The effects of QP poisoning appear as random telegraph noise in the data and can be modeled as a Lorentzian~\cite{brock2021nonlinear}. We thus employ a combined model including a Lorentzian and a power law fit to describe the measured apparent charge noise $S^{\mathrm{meas}}_{qq} (\omega)=S_{\mathrm{QP}}$+$S_{qq}^{\mathrm{int}}$. However, the roll-off frequency for the Lorentzian fit is not resolvable using this measurement, as the bandwidth of the fit is limited by the lock-in roll-off frequency 1331 Hz. Moreover, the accuracy decreases for frequencies $>$ 200 Hz where the SNR $\sim$ 1. Hence, the noise floor due to QP appears to be white noise rather than Lorentzian. As the contributions to this offset-noise were observed to decrease for lower $n_g$ values where the effects of quasiparticles are also reduced, we believe our Lorentzian model holds validity. Fig. \ref{fig:fig_ccpt_exp}(d) displays the calculated $S_{qq}^{\mathrm{int}}(\omega)$ varying as
\begin{equation}
    S_{qq}^{\mathrm{int}}(\omega) = (5.5 \times 10^{-7}) \left(\frac{\omega}{2\pi}\right)^{-0.89} \mathrm{e}^2/\mathrm{Hz}.
\end{equation}
The total standard deviation of charge fluctuations calculated over the bandwidth 1 Hz to $\omega_{\mathrm{LPF}}/2\pi$ Hz is found to be 2.5 $\times 10^{-3}$ electrons. This value aligns with previously reported measurements of charge fluctuations for this device to within an order of magnitude~\cite{brock2021nonlinear}. 

As mentioned in Sec. \ref{concept}, ideally we prefer $K(\omega) \gg G(\omega)^{-1}$ such that $Y_\mathrm{m}(t)$ follows $Y_{\mathrm{ref}}$ closely. However, this is accompanied by an increase in the pick-up of the noise floor as well~\cite{bechhoefer2005feedback}. We may balance out the combined effects of faster noise suppression and increased sensor-noise pick-up by shaping the net loop gain to follow $T(\omega) = (1+i\omega/\omega')^{-1}$, where $\omega'$ is the feedback bandwidth. This can be accomplished using proportional-integral control, by fixing $K(\omega) = \omega'(1+\omega_{\mathrm{LPF}}/\omega)/G_0\omega_{\mathrm{LPF}}$. We can furthermore choose $\omega'$ such that $K(\omega)G(\omega) = \omega'/\omega \gg 1$ in the region where we have an appreciable SNR [refer to Fig. \ref{fig:fig_ccpt_exp}(c)], but drops later as the SNR plunges.
\section{Results and discussion}
\label{results}
\begin{figure*}[thb]
\includegraphics[height=0.2\textheight, trim = 0 0 0 0]{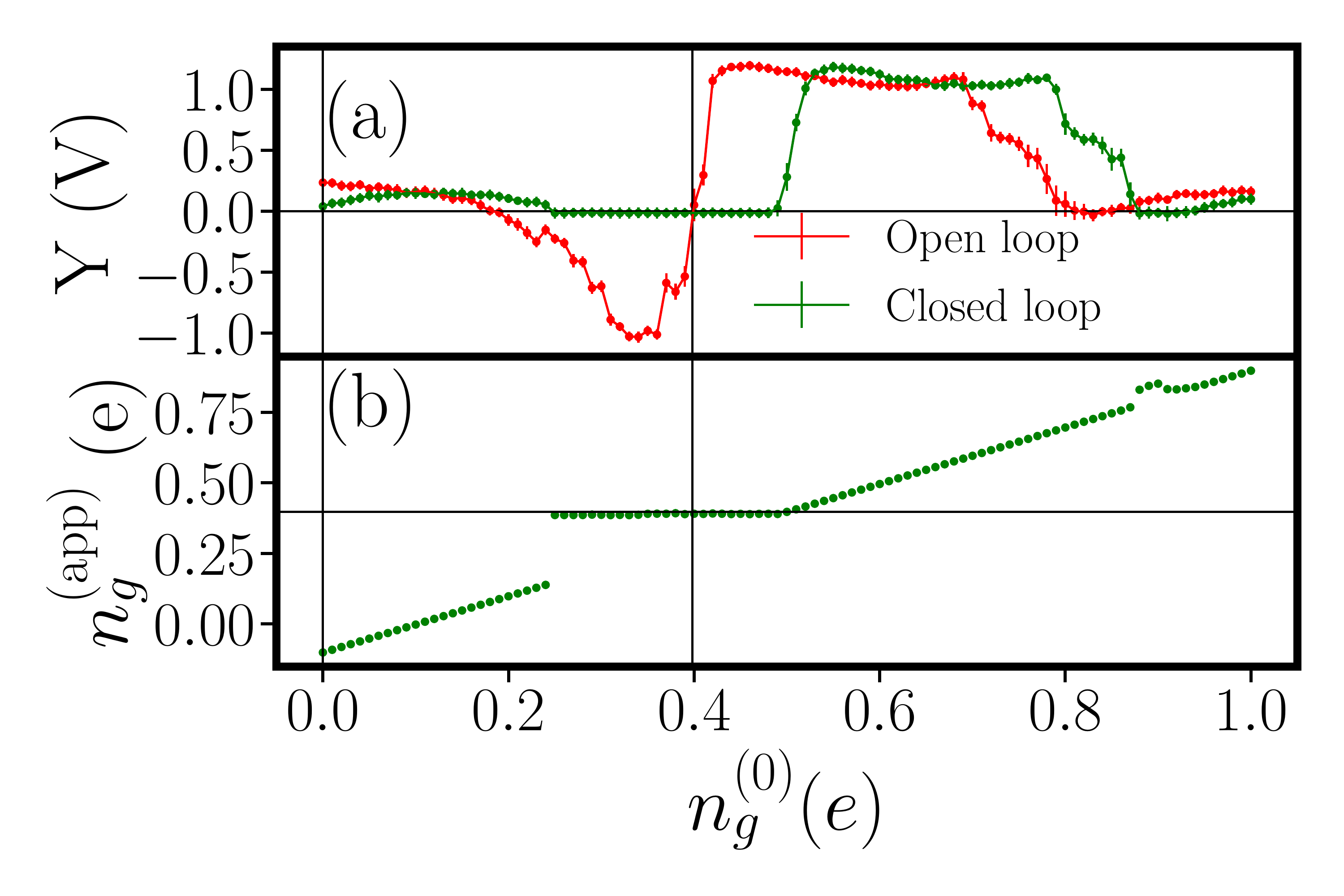}
\hspace{0.5cm}
\includegraphics[height=0.2\textheight, trim = 0 0 0 0]{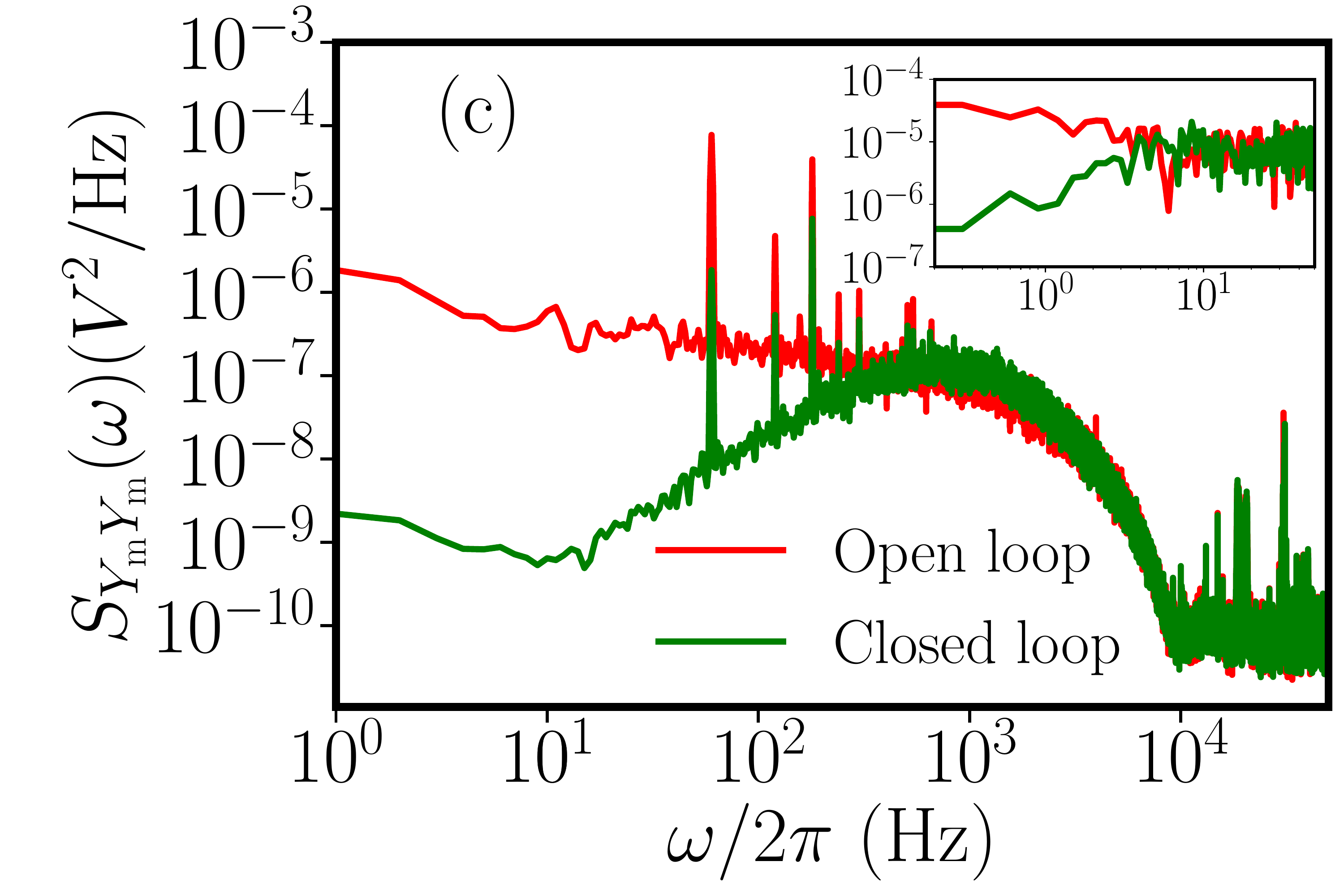}\\
\vspace{0.2cm}
\includegraphics[height=0.2\textheight, trim = 80 0 30 30]{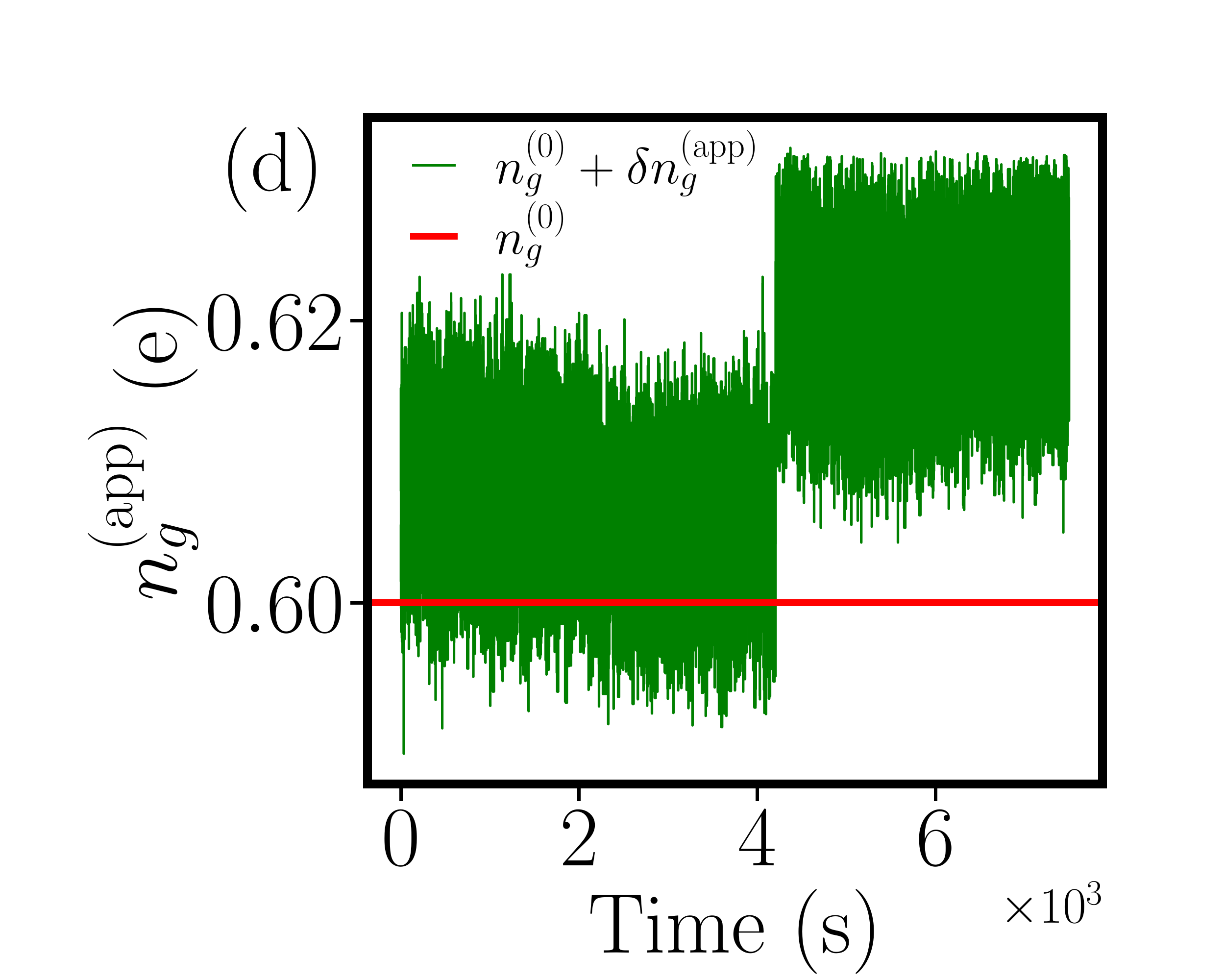}
\includegraphics[height=0.2\textheight, trim = 0 0 0 30]{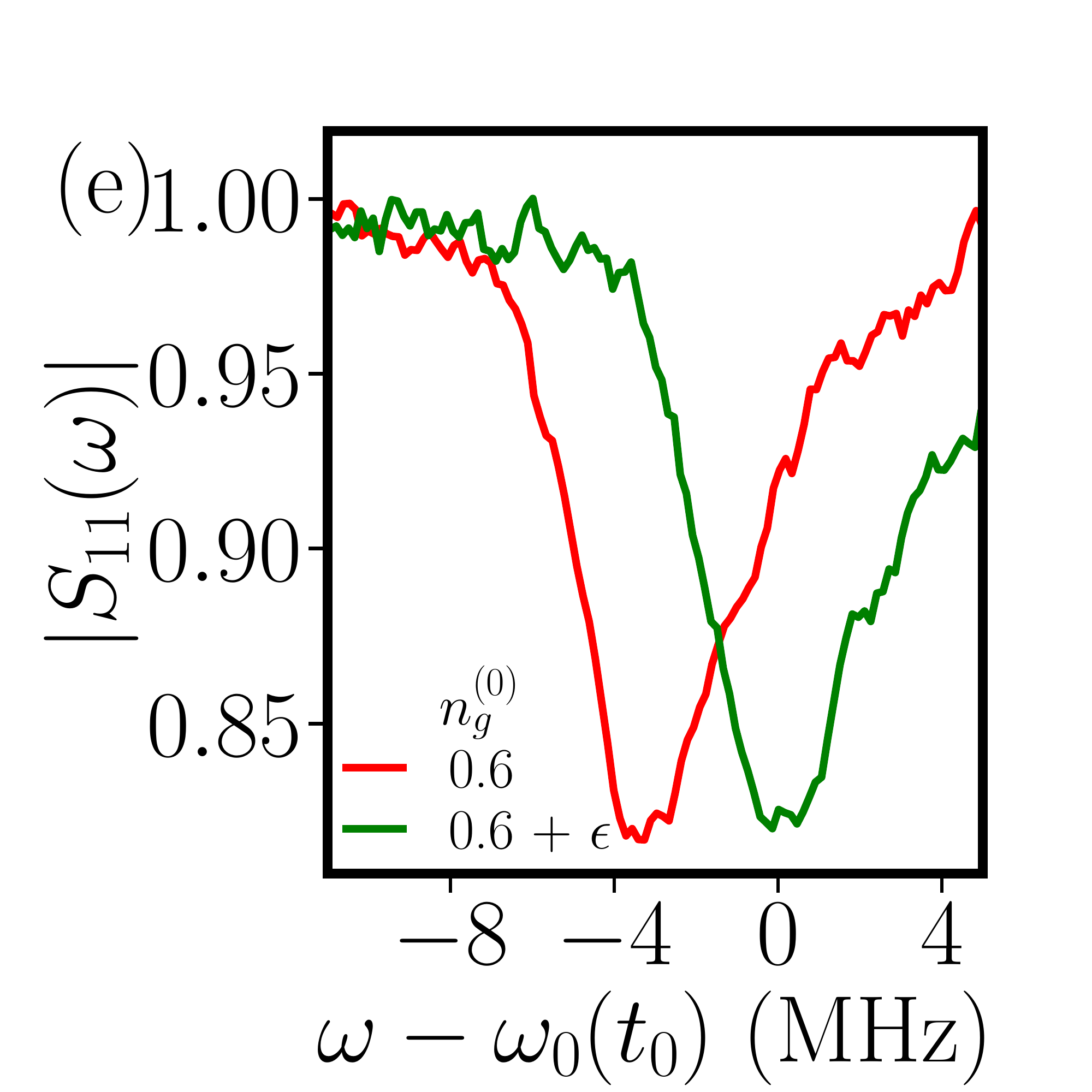}
\includegraphics[height=0.2\textheight, trim = 0 0 0 0]{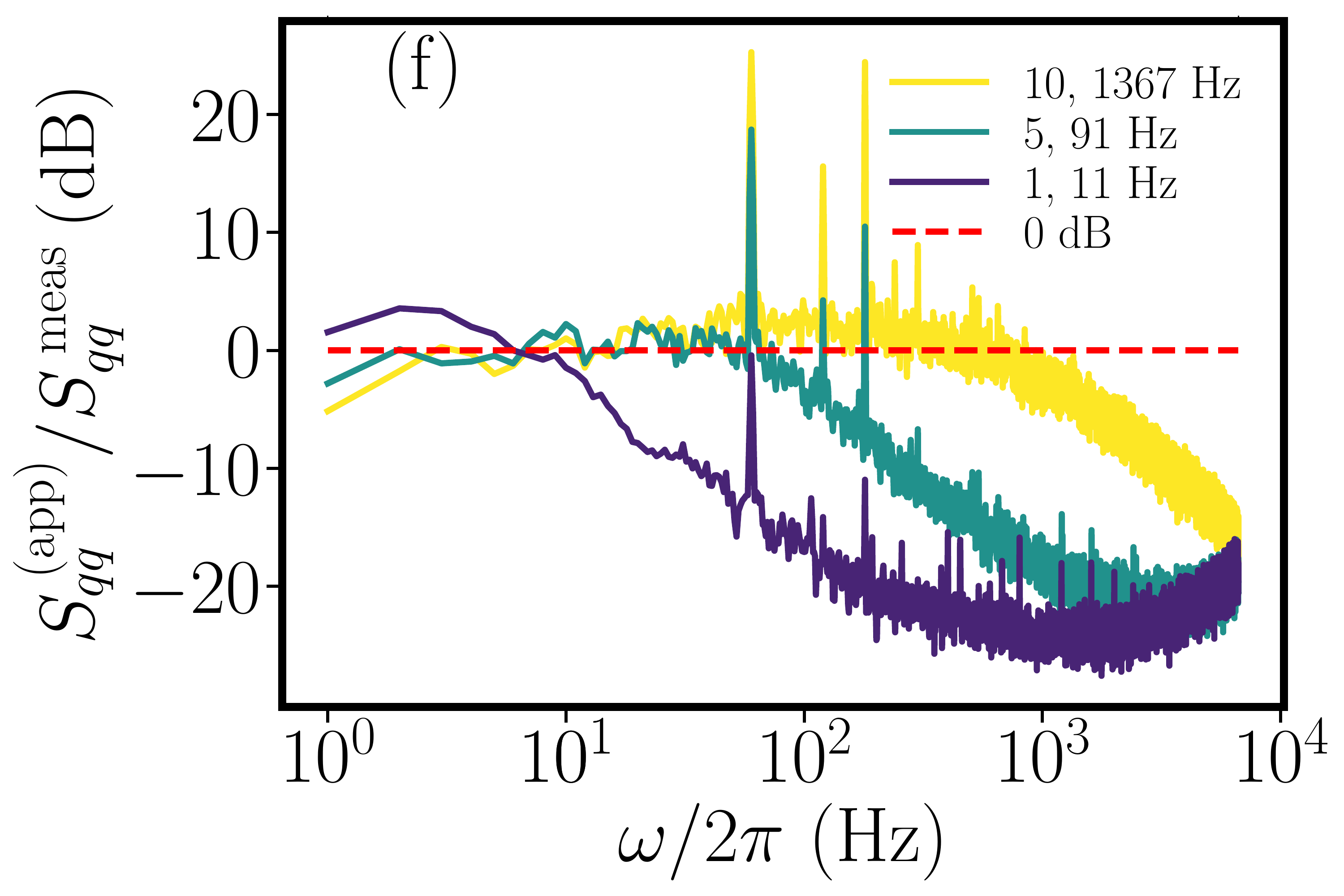}
\caption{(a) Proof of concept for charge noise correction under feedback locking. The cCPT is gate-swept from $0\leq n_g^{(0)}\leq1$, for $\omega_c = \omega_0(0.4,0)$ and $n=5$. Each point corresponds to the averaged value of the measurements spanning 1 sec with a sampling rate 1 kHz, and the time constant set to 10 msec. Error bars are also observed to decrease once the feedback is locked. For e.g., at $n_g^{(0)}=0.4$, the standard deviations of the measured data are 135 mV and 50 mV for open-loop and closed-loop configurations, respectively. (b) Net corrected $n_{g}^{(\mathrm{app})} = n_g^{(0)} + \delta n_{g}^{(\mathrm{app})} (t)$ for the data in (a). Due to the PID correction, $n_{g}^{(\mathrm{app})}$ is set to 0.4, roughly across the region $|\delta n_g| \equiv |n_g^{(0)} - 0.4| \leq 0.1$. (c) Comparison of measured $S_{Y_\mathrm{m}Y_\mathrm{m}}(\omega)$ displays a definitive suppression in the resonant fluctuations at $n=10$ and bias point (0.6,0). The inset displays the comparison of $S_{Y_\mathrm{m}Y_\mathrm{m}}(\omega)$ at $n=1$ and bias point (0.4,0). Note the measurements for $n=10$ and $n=1$ are taken for different values of gains of the lock-in amplifier. (d) Time domain data of $n_g^{(\mathrm{app})}$ during the feedback operation. The measurement is completed in $7.5 \times 10^3$ sec with a sampling rate of 10 Hz. The cCPT is biased at (0.6,0) and the cavity is driven at $n=10$. The red plot displays $n_g^{(0)} = 0.6$ and the green plot displays the net, feedback corrected response. As can be seen in this plot, the CPT underwent a discrete jump in the gate offset at the island during the course of this measurement. This can be considered as a quasi-static event as it occurs at a very low frequency. (e) Reflection coefficients of the cavity taken after the measurement in (d). Red and green plots correspond to the cCPT biased at $n_g^{(0)}=0.6$ (red), and $n_g^{(0)}=0.6+\epsilon =0.622$ (green), respectively. Here, $ \epsilon\equiv \langle \delta n_g^{\mathrm{(app)}}(t) \rangle$ and is averaged over the last bins of data in Fig. \ref{fig:fig_ccpt_res}(d). Due to the discrete jump in gate charge, the resonant frequency shifted nearly 4 MHz but the feedback configuration accurately tracks this event. (f) $S_{qq}^{(\mathrm{app})}(\omega)/S_{qq}^{\mathrm{meas}}(\omega)$ in units of dB for different photon numbers $n$. The dashed red plot is the fit obtained from Fig. \ref{fig:fig_ccpt_exp}(d) and corresponds to the measured apparent charge noise, to act as a reference. The legends display $n$ and the calculated 3 dB roll-off frequency for the corresponding plot. Except for $n=10$, the rest of the measurements were taken with the lock-in amplifier time constant set at 300 $\mu$sec. The cCPT is biased at (0.6, 0) for $n=10$ and $n=5$, and at (0.4, 0) for $n=1$.}
\label{fig:fig_ccpt_res}
\end{figure*}
The feedback correction for the charge noise is measured via the simultaneous detection of $Y_\mathrm{m}(t)$ and the input gate correction, by means of a digitizer. Figures \ref{fig:fig_ccpt_res}(a) and \ref{fig:fig_ccpt_res}(b) provide proof of concept for our scheme. Both $Y_\mathrm{m}$ [refer to \ref{fig:fig_ccpt_res}(a)] and the total averaged gate charge including the PID correction, i.e., $n_{g}^{(\mathrm{app})} = n_g^{(0)} + \delta n_{g}^{(\mathrm{app})} (t)$ [refer to \ref{fig:fig_ccpt_res}(b)], are measured as the cCPT is gate-swept from $0\leq n_g^{(0)}\leq1$, for $\omega_\mathrm{c} = \omega_0(0.4,0)$. The quadrature $Y_\mathrm{m}$ is nulled, and $n_{g}^{(\mathrm{app})}$ is set to 0.4, roughly across the region $|\delta n_g| \equiv |n_g^{(0)} - 0.4| \leq 0.1$. Note that the feedback correction continues in the right direction as long as $\text{sgn}(\delta n_g) = \text{sgn}(Y_\mathrm{m})$, until $Y_\mathrm{m}$ changes sign; hence the corrected bandwidth applies to $\delta \omega_0>\kappa_{\mathrm{tot}}$ as well, and the feedback, once locked, is robust against discrete gate-jumps of small magnitude.

The reduction of resonant frequency fluctuations can be directly observed by comparing the open and closed loop PSDs for $Y_\mathrm{m}(t)$. This is shown in Fig. \ref{fig:fig_ccpt_res}(c) and is measured under the same configuration as discussed in Figs. \ref{fig:fig_ccpt_exp}(c) and \ref{fig:fig_ccpt_exp}(d). Note that the detected 60, 120 and 180 Hz peaks are primarily from the compressors and pumps feeding our cryostat, and are sources of external noise. The inset displays the comparison of $S_{Y_\mathrm{m}Y_\mathrm{m}}(\omega)$ at $n=1$ and bias point (0.4,0).

In order to test the durability of the feedback loop, the system was monitored for $7.5 \times 10^3$ sec ($\sim$2 hours), with $n_g^{(0)}$ chosen as 0.6 and the flux at a minimally sensitive point, and with the cavity driven at $n=10$. 
Figures \ref{fig:fig_ccpt_res}(d) and (e) demonstrate the efficiency of the closed-loop system during the event of a discrete jump in gate charge, as mentioned before. Figure \ref{fig:fig_ccpt_res}(d) displays the time-domain data of $n_g^{\mathrm{(app)}}$ collected with a sampling rate of 20 Hz. As can be seen in this plot, the CPT underwent a discrete jump in the gate offset at the island during the course of this measurement. Figure \ref{fig:fig_ccpt_res}(e) plots the reflection coefficient $|S_{11}(\omega)|$ immediately after the measurement in \ref{fig:fig_ccpt_res}(d). Here, $n_g^{(0)} = 0.6$ corresponds to the unlocked value (red) and $n_g^{(0)} = 0.6+ \epsilon$ corresponds to the feedback-locked value (green), where $ \epsilon\equiv \langle \delta n_g^{\mathrm{(app)}}(t) \rangle$, averaged over the last bins of data in Fig. \ref{fig:fig_ccpt_res}(d). As can be seen, the resonance underwent a shift of nearly 4 MHz due to the gate-charge jump during the measurement, and gets accurately tracked by the loop.
It is to be noted that longer measurements also undergo a slow drift in the internal bias noise due to the presence of low-frequency components. As a result, $Y_\mathrm{m}(t)$ deviates from the linear response described in Eq.(\ref{yquadeqn}), and becomes second-order, picking up contributions from $\delta b_{\mathrm{int}}^2(t)$. The PSD of the charge noise extraction from $S_{\mathrm{Y_\mathrm{m}Y_\mathrm{m}}}(\omega)$ as described in Eq. (\ref{syy_eqn}) breaks down in this regime. 

Finally, Fig. \ref{fig:fig_ccpt_res}(f) captures the feedback response for varying photon numbers $n$ = 10, 5 and 1, by plotting the PSD of the applied gate charge, $S_{qq}^{(\mathrm{app})}(\omega)$, in comparison to the measured apparent charge noise $S^{\mathrm{meas}}_{qq}$. The dashed red plot is shown for reference, and represent the 0 dB point. We cannot accurately extract the noise floor in the closed loop setup since the gain of the transfer function changes. However, by placing $S^{\mathrm{meas}}_{qq}(\omega)$ as a reference, we can ensure that the net corrected gate PSD does not over-compensate for the noise floor fluctuations. This is important because of the smallness of the SNR, especially at $n=1$. As can be observed, at $n=10$, $n_g^{(\mathrm{app})}(\omega)$ follows the measured apparent charge noise closely. This implies a significant correction for the intrinsic charge noise, and the stabilization of the resonant frequency, with a roll-off set by the 3 dB point at $\sim$1.4 kHz. Due to the decrease in SNR($\omega$) as $n$ is lowered [refer to Eq.(\ref{G0eqn})], we use a longer time constant for $n$=5 through $n$=1, resulting in a significant decrease in roll-off frequency near the single-photon limit.

Note that since the chosen bias point for $n$=10 and 5 in Fig. \ref{fig:fig_ccpt_res}(f) is $(n_g, \Phi_{\mathrm{ext}}) = (0.6, 0)$, the net applied gate charge also accounts for the QP switching noise. In contrast, the bias point is fixed at $(n_g, \Phi_{\mathrm{ext}})=(0.4,0)$ for the single-photon case. The resulting feedback response better tracks the actual intrinsic noise in this regime since the QP interference is significantly reduced.

We observe that a major limitation in the efficient correction for charge noise at single photon occupancy of the cavity is the drastic decrease in SNR($\omega$). Along with the noise contributions from the amplifier chain at the TWPA, HEMT, and FET stages, the power detector amplifies the noise floor correlations at $\omega_\mathrm{m}$ over the tunable bandpass filter bandwidth of 80 MHz. This can be best circumvented by using a series of notch filters before the detector with effective stop-bands within the 80 MHz bandwidth of the band-pass filter, but with pass-bands at $\omega_\mathrm{c}$ and $\omega_\mathrm{c} \pm \omega_\mathrm{m}$. This ensures that detector input consists of mostly signal frequencies, thus decreasing the noise floor of the transfer function [refer to Fig. \ref{fig:fig_ccpt_exp}(c)].

As is evident in the previous discussion, another limiting constraint in our setup is the existence of quasiparticle poisoning in the CPT. This affects our choice of parameters in three ways. Firstly, the probability of switching to the odd electron state increases steadily towards charge degeneracy, due to its more favorable electrostatic energy configuration as compared to CPT's even band~\cite{aumentado-nonequilibrium-2004,lutchyn-effect-2007}. The effect of quasiparticles on the extraction of the error signal can be observed in Fig. \ref{fig:fig_ccpt_res}(a) near $n_g$ = 0.8, where the resonance has completely switched to the odd parity. We therefore avoid operating the feedback at $|n_g| \geq 0.65$ to evade accidental destabilization of the loop. Moreover, near $n_g$ = 0.5, $\delta \omega_0^{\mathrm{qp}} = |\delta \omega_0^{(\mathrm{odd})} - \delta \omega_0^{(\mathrm{even})}| < \kappa_{\mathrm{tot}}$. This can smear out the smooth monotonic function preferred for the accurate detection of charge noise using $Y_\mathrm{m}(t)$. Finally, $\omega_\mathrm{m}$ is chosen such that the sidebands are ensured to be away from both of the resonant frequencies. If $|\omega_\mathrm{m} - \delta \omega_0^{\mathrm{qp}}|< \kappa_{\mathrm{tot}}$, this assumption does not hold and results in a non-zero $|Y(n_g)|$ at resonance. In other cases, the sine quadrature is expected to detect a null signal whenever the cavity switches out of resonance (typically at frequencies 1 kHz - 100 kHz) and the effects of QP can be accounted for empirically as discussed in Sec. \ref{experiment}~\cite{aumentado2010cooper}.  

The demonstration of charge noise correction reported in this work can also, in principle, be extended to reduce the effects of flux noise in the cCPT. However, in our setup, the DC flux line undergoes heavy filtering (with a cut-off frequency of 10 Hz) due to the RC low pass filter formed by the current limiting resistor and capacitor. The parasitic capacitance in the gate line leads to a RC filtering with cut-offf frequency $>$400 kHz. This ensures the feedback correction is not affected by the transfer function of the gate line itself, as opposed to the flux source.
\section{Conclusion}
\label{conclusion}
In this work, we successfully demonstrate feedback stabilization of a tunable microwave cavity against intrinsic charge noise by locking the cavity to a stable reference. We report stabilization of the cavity resonance over a 3dB bandwidth of 1.4 kHz at $n=10$. When the cavity is driven at the single photon level, this bandwidth is reduced to 11 Hz, due to the accompanying decrease in SNR. Compensation for intrinsic bias noise stabilizes the resonant frequency with respect to the carrier signal over the course of an actual measurement, as in electrometry and qubit readout. We believe that the resulting enhancement in charge sensitivity can raise the cCPT's performance to operate in the regime of single photon-phonon coupled optomechanics. The feedback scheme reported here can also be extended to tunable microwave cavities in general, provided the dominant source of resonant frequency fluctuations originate from the intrinsic bias noise at the sample. The technique can thus realize real-time detection and correction for bias noise in these devices, potentially improving the coherence and measurement fidelities in superconducting qubits.
\begin{acknowledgments}
We thank Andrew D. Armour and William Braasch for very helpful discussions. This work was supported by the NSF under Grants No. DMR-1807785 (S. K., B. T., B. L. B., and A. R) and  DMR-1507383 (M. P. B.), and by a Google research award (S. K.). S. K.
also acknowledges the support of a Gordon F. Hull
Dartmouth graduate fellowship. 
\end{acknowledgments}

\section*{Data Availability Statement}
The data that support the findings of this study are available from the corresponding author upon reasonable request.

\end{document}